\documentclass[10pt,twocolumn,twoside]{IEEEtran} 
\usepackage{cite}
\usepackage{amsmath,amssymb,amsfonts}
\usepackage{algorithmic}
\usepackage{graphicx}
\usepackage{textcomp}
\usepackage{multirow}
\usepackage{siunitx}
\usepackage{mathtools}
\usepackage{romannum}
\usepackage{kotex}
\usepackage{algorithmic}
\usepackage{algorithm}
\usepackage{enumitem}
\usepackage{xcolor}
\usepackage{threeparttable}
\usepackage[noabbrev]{cleveref}
\crefrangeformat{equation}{(#3#1#4)-(#5#2#6)}
\usepackage{cite}
\usepackage{tikz}   
\usepackage{tikz-qtree}
\usepackage{comment}
\usetikzlibrary{trees,calc,arrows.meta,positioning,decorations.pathreplacing,bending}
\usetikzlibrary{shapes,arrows}
\tikzset{
    edge from parent/.style={draw, thick,black},
    no edge from this parent/.style={
        every child/.append style={
        edge from parent/.style={draw=none}}}
         }
\usepackage{pgfplots}
\usepackage{grffile}
\pgfplotsset{compat=newest}
\usetikzlibrary{plotmarks}
\usetikzlibrary{arrows.meta}
\usepgfplotslibrary{patchplots}

\newcommand{\xvdots}{\vphantom{{\sum^0}^0}\smash{\vdots}}

\newcommand{\rmm}{{\mathrm{m}}}
\newcommand{\rmM}{{\mathrm{M}}}
\newcommand{\rme}{{\mathrm{e}}}
\newcommand{\rma}{{\mathrm{a}}}
\newcommand{\rmb}{{\mathrm{b}}}
\newcommand{\rmc}{{\mathrm{c}}}
\newcommand{\rmK}{{\mathrm{K}}}
\newcommand{\rmk}{{\mathrm{k}}}
\newcommand{\rmu}{{\mathrm{u}}}
\newcommand{\rmx}{{\mathrm{x}}}
\newcommand{\rmA}{{\mathrm{A}}}

\newcommand{\rrr}{{\mathsf{r}}}
\newcommand{\sss}{{\mathsf{s}}}
\newcommand{\LLL}{{\mathsf{L}}}

\newcommand{\calO}{{\mathcal{O}}}

\newcommand{\cc}{{\mathbf{c}}}
\newcommand{\CC}{{\mathbf{C}}}
\newcommand{\xx}{{\mathbf{x}}}
\newcommand{\uu}{{\mathbf{u}}}
\newcommand{\yy}{{\mathbf{y}}}
\newcommand{\FF}{{\mathbf{F}}}
\newcommand{\HH}{{\mathbf{H}}}
\newcommand{\GG}{{\mathbf{G}}}

\newcommand{\Z}{\ensuremath{{\mathbb Z}}}
\newcommand{\N}{\ensuremath{{\mathbb N}}}
\newcommand{\R}{\ensuremath{{\mathbb R}}}

\newcommand{\Q}{\ensuremath{{\mathbb Q}}}

\newcommand{\idx}{{\mathsf{idx}}}
\newcommand{\BRidx}{{\mathsf{BRidx}}}
\newcommand{\Enc}{{\mathsf{Enc}}}
\newcommand{\Decomp}{{\mathsf{D}}}
\newcommand{\Slot}{{\mathsf{Slot}}}
\newcommand{\nom}{{\mathsf{nom}}}
\newcommand{\sk}{{\mathrm{sk}}}
\newcommand{\ak}{{\mathrm{ak}}}
\newcommand{\Dec}{\mathsf{Dec}}

\newcommand{\Mult}{\mathsf{Mult}}
\newcommand{\ini}{\mathsf{ini}}
\newcommand{\Pack}{\mathsf{Pack}}

\newcommand{\UnpackPt}{\mathsf{UnpackPt}}
\newcommand{\UnpackCt}{\mathsf{UnpackCt}}
\newcommand{\Bitreverse}{\mathsf{BitReverse}}

\newcommand{\Ceil}{\ensuremath{\left\lceil}}

\newcommand{\Flr}{\ensuremath{\right\rfloor}}

\newcommand{\ra}{\ensuremath{\rightarrow}}

\newtheorem{thm1}{\bf Theorem}
\newtheorem{prop1}{\bf Proposition}
\newtheorem{lem1}{\bf Lemma}

\newtheorem{defn1}{\bf Definition}
\newtheorem{rem1}{\bf Remark}
\newtheorem{exmp1}{\bf Example}
\newtheorem{cor1}{\bf Corollary}

\newtheorem{prob1}{\bf Problem}

\renewcommand{\algorithmicrequire}{\textbf{Input:}}

\def\BibTeX{{\rm B\kern-.05em{\sc i\kern-.025em b}\kern-.08em
    T\kern-.1667em\lower.7ex\hbox{E}\kern-.125emX}}

\begin{document}
\title{Ring-LWE based Encrypted Controller with Unlimited Number of Recursive Multiplications and  Effect of Error Growth}

\author{Yeongjun Jang,
Joowon Lee, 
Seonhong Min, 
Hyesun Kwak,
Junsoo Kim, and
Yongsoo Song
\thanks{This work was supported by the National Research Foundation of Korea(NRF) grant funded by the Korea government(MSIT) (No. RS-2022-00165417 and No. RS-2024-00353032).
} 
\thanks{Y.~Jang and J.~Lee are with ASRI, the Department of Electrical and Computer Engineering, Seoul National University, Seoul, 08826, Korea (email: \{jangyj, jwlee\}@cdsl.kr).
} 
\thanks{S.~Min, H.~Kwak, and Y.~Song are with the Department of Computer Science and Engineering, Seoul National University, Seoul, 08826, Korea (email: \{hskwak, minsh, y.song\}@snu.ac.kr).
}
\thanks{J.~Kim is with the Department of Electrical and Information Engineering, Seoul National University of Science and Technology, Seoul, 01811, Korea (email: junsookim@seoultech.ac.kr).
}
}

\maketitle
\begin{abstract}
In this paper, we propose an encrypted dynamic controller that executes an unlimited number of recursive homomorphic multiplications on a Ring Learning With Errors (Ring-LWE) based cryptosystem without bootstrapping.
The proposed controller exhibits lower computational complexity compared to existing encrypted controllers implemented on LWE based schemes due to the polynomial structure of Ring-LWE.
However, the structural difference introduces additional difficulties in analyzing the effect of error growth; Ring-LWE based schemes inject multiple error coefficients when encrypting a single message,
which accumulate under recursive homomorphic multiplications.   
We show that their effect on the control performance can be arbitrarily bounded by the closed-loop stability, thus recovering the performance of the unencrypted controller. 
Furthermore, a novel method to ``pack’' a vector into a polynomial is presented, which enhances computational and memory efficiency when applied to the proposed encrypted controller.
The effectiveness of the proposed design is demonstrated through numerical simulations.
\end{abstract}

\section{Introduction}\label{Sec:Intro}

There is a growing emphasis on securing networked control systems from cyber attacks that could cause significant physical damage 
\cite{Sandberg 15}.
The urgency of this matter has been amplified by the recent development of attack methodologies that exploit disclosed control signals and/or model knowledge to avoid detection 
\cite{Teixeira 15}.
One of the initial countermeasure was to encrypt the control signals before transmission, which are then decrypted at the controller for computation.
However, since the controller resides at the cyber layer in networked control systems, the decrypted signals and the parameters of the controller remain vulnerable to adversaries.
In this context, the notion of encrypted control \cite{Kogiso 15,Schulze Darup 21,ARC} has been introduced.
Encrypted controllers utilize homomorphic encryption (HE) to operate directly over encrypted data without decryption, thereby enabling all control signals and parameters to remain encrypted throughout the cyber-layer.

When running dynamic controllers over encrypted data, it is essential to account for effect of error growth caused by the recursive update of the (encrypted) state. 
This is because most HE schemes that support both addition and multiplication over encrypted data are based on the Learning With Errors (LWE) problem \cite{Regev 09}. 
LWE based schemes inject random errors during encryption to ensure security that could accumulate and contaminate the message part under recursive operations.
The bootstrapping technique \cite{Gentry 09} could be employed to refresh these errors, but it is not yet considered practical for real-time control systems.

The effect of error growth has been addressed from a control-theoretic perspective in \cite{Kim 23}. 
It has been shown that the closed-loop stability of the plant and the
controller can suppress the effect of error growth on the control performance, even though the encrypted controller performs recursive multiplications on a LWE based scheme. 
However, the multiplication technique \cite{Gentry 13} used in \cite{Kim 23} demands high computational resources and a large amount of memory usage.

In fact, there is an algebraic variant of LWE, called Ring-LWE \cite{Lyubashevsky 13}, which is more efficient in terms of computation and memory, and offers the same level of security.
In this regard, widely used HE libraries such as Microsoft SEAL \cite{Microsoft Seal}, HElib \cite{HElib}, and Lattigo \cite{Lattigo} are built upon Ring-LWE based cryptosystems.
Such efficiency comes from their structure based on polynomial rings; however, this also introduces a further challenge regarding the effect of error growth.
Whereas LWE based schemes inject errors as scalars, Ring-LWE based schemes inject errors as polynomials, where each coefficient is a scalar error.
These error coefficients accumulate under recursive multiplications (see \eqref{eq:decoupleHigh}).

Instead of analyzing the effect of such error growth, recent studies \cite{Teranishi23,Lee 24} on utilizing Ring-LWE based cryptosystems to encrypt dynamic controllers avoid recursive multiplications with the help of re-encryption, which requires an extra communication line between the plant and the controller.
To the best of our knowledge, \textit{a method to perform an unlimited number of recursive homomorphic multiplications on a Ring-LWE based cryptosystem without bootstrapping has not yet been reported in literature.}

In this paper, we propose an encrypted dynamic controller that executes an unlimited number of recursive multiplications on a Ring-LWE based cryptosystem without bootstrapping.
In the proposed design, only the error coefficients in the constant terms affect the control performance. 
It is shown that their effects can be arbitrarily bounded by means of the closed-loop stability, even when the \textit{non-effective} error coefficients in the non-constant terms overflow the bounded message space of the Ring-LWE based scheme.
For multiplications, we employ the external product of \cite{Chillotti 16}, called the ``Ring-GSW (Gentry-Sahai-Waters) scheme,''
which offers lower computational complexity than the multiplication technique on LWE based schemes used in \cite{Kim 23}.
It is also worth noting that the proposed controller's ability to perform recursive multiplications obviates the need of re-encryption, in order to refresh grown errors.

Furthermore, we propose and apply
a novel method to encode a vector into a polynomial---such method is called ``packing'' in cryptography---that accelerates computation and improves memory efficiency.
The proposed packing algorithm assigns each element of a vector as the coefficient to the designated term of a polynomial.
This enables multiple scalar operations to be replaced by a single polynomial operation.
We present an encrypted controller equipped with the packing algorithm and again analyze the effect of error growth, which can also be arbitrarily bounded.
The efficiency of utilizing the packing algorithm is analyzed and demonstrated through numerical simulations.

The remainder of the paper is organized as follows. 
Section~\ref{Sec:Preliminary} introduces a Ring-LWE based scheme and formulates the problem. 
Section~\ref{Sec:Naive} presents the proposed encrypted dynamic controller, and Section~\ref{Sec:Hoisted} describes and applies the proposed packing algorithm to the encrypted controller. 
Section~\ref{Sec:Simulation} provides the simulation results. 
Finally, Section~\ref{Sec:Conclusion} concludes the paper.

\indent \textit{Notation:} The set of integers, non-negative integers, positive integers, rational numbers, and real numbers are denoted by $\mathbb{Z}$, $\mathbb{Z}_{\ge 0}$, $\mathbb{N}$, $\mathbb{Q}$, and $\R$ respectively. 
The floor, ceiling, and rounding operations are denoted by $\lfloor \cdot \rfloor$, $\lceil \cdot \rceil$, and $\lceil \cdot \rfloor$, respectively. 
For $q\in\mathbb{N}$, we define $\mathbb{Z}_q$ by the set of integers in the interval $[-q/2, q/2)$. 
For $a\in\mathbb{Z}$ and $q\in\mathbb{N}$, we define the modulo operation by $a \!\!\mod q := a - \lfloor (a+q/2)/q\rfloor q$. 
The floor, ceiling, rounding, and modulo operations are defined component-wisely for vectors and matrices, and coefficient-wisely for polynomials.
The Kronecker product is denoted by $\otimes$.
For a (matrix) vector of scalars, $\| \cdot\|$ denotes the (induced) infinity norm.
For a sequence $v_1,\dots,v_n$ of scalars or column vectors, we define $[v_1;\cdots;v_n ]:=[v_1^\top \cdots v_n^\top ]^\top$.
For a polynomial $\rma(X) = \sum_{i=0}^{N-1}\rma_iX^{i}$, the $l_\infty$ norm is denoted by $\|\rma(X) \|:= \max_{0\le i < N}\{  | \rma_i| \}$. 
For a matrix of polynomials with $m\in\N$ rows and $n\in\N$ columns denoted by $\rmA(X)=\{\rmA_{i,j}(X)\}$, we define $\|\rmA(X)\|:= \max_{1\le i\le m}\sum_{j=1}^n\|\rmA_{i,j}(X) \|$, and $\rmA(X)\big|_{X=0}:=\{\rmA_{i,j}(0)\}$.
For $m\in\N$ and $n\in \N$, let $0_{m\times n}\in\Z^{m \times n}$ and $I_n \in \Z^{n \times n}$ denote the zero matrix and the identity matrix, respectively.

\section{Preliminaries and Problem Formulation}\label{Sec:Preliminary}

\subsection{Ring-LWE based cryptosystem}\label{subsec:rlwe}
This subsection briefly explains the Ring-LWE based cryptosystem \cite{Lyubashevsky 13,Chillotti 16} used in this paper.
The scheme involves two encryption algorithms, namely the Ring-LWE encryption and the Ring-GSW encryption. 
We will describe each encryption algorithm and illustrate how to perform
additions and multiplications over encrypted data.

To describe the scheme, the following definitions are introduced.
For $N\in\N$, let $R :=\mathbb{Z}[X] / \langle X^N+1 \rangle $ be the ring of integer polynomials modulo $X^N+1$, whose elements are represented by integer polynomials of degree less than $N$. 
For $q\in\N$, let $R_q := \mathbb{Z}_q[X ] / \langle X^N+1 \rangle$ be the residue ring of $R$ modulo $q$.
Every element of the ring $R_q$ is represented by a polynomial of degree less than $N$, whose coefficients belong to the set $\Z_q$.

The addition and multiplication of polynomials $\rma(X) = \sum_{i=0}^{N-1}\rma_iX^{i}\in R_q$ and $\rmb(X) = \sum_{i=0}^{N-1}\rmb_iX^{i}\in R_q$ are defined by 
\begin{equation}\label{eq:defAddMult}
    \begin{split}
        \rma(X) + \rmb(X) &:= \textstyle\sum_{i=0}^{N-1} \left(\rma_i+\rmb_i\right)X^{i} \!\!\mod q, \\
        \rma(X)\cdot \rmb(X) &:=\textstyle\sum_{i=0}^{N-1}\rmc_iX^{i} \!\!\mod q,
    \end{split}    
\end{equation}
where $\rmc_i:=\sum_{j=0}^{i}\rma_j\rmb_{i-j}-\sum_{j=i+1}^{N-1}\rma_{j}\rmb_{N+i-j}$. 
These operations can be interpreted as general addition and multiplication of real polynomials, where $X^N$ is regarded as $-1$ and the coefficients are mapped to the set $\mathbb{Z}_q$ via the modulo operation.  
Although the modulo operation is embedded in the definitions given in \eqref{eq:defAddMult}, we often deliberately indicate it and write $\rma(X)+\rmb(X)\!\!\mod q$ or $\rma(X)\cdot \rmb(X)\!\!\mod q$ for readers who are not familiar with operations on polynomial rings.

Now we describe 
the Ring-LWE based scheme \cite{Lyubashevsky 13}.
Ciphertexts (encrypted data) are written in bold type to distinguish them from plaintexts (messages to be encrypted).
 \begin{itemize}[leftmargin=*]
        \item ($\mathtt{Setup}$) Choose a power of two $N\in\mathbb{N}$ and a prime $q\in\mathbb{N}$ such that $1=2N \!\! \mod q$.
        Fix an error distribution $\psi$ over $R$ bounded by $\sigma >0$, i.e., any $\rme\in R$ sampled according to $\psi$ can be denoted by 
         \begin{equation}\label{eq:RLWEerr}
            \rme = \rme_0+\rme_1X+\cdots+\rme_{N-1}X^{N-1}
        \end{equation}
        where $| \rme_i |\le\sigma$ for $i=0,1,\dots,N-1$.
        \item ($\mathtt{Key\ generation}$) The secret key $\sk\in R$ is sampled according to $\psi$.
        \item ($\mathtt{Ring\text{-}LWE \ encryption}$) $\Enc: R_q \rightarrow R_q^2$. Sample a polynomial $\rma \in R_q$ uniformly at random from $R_q$, and an error polynomial $\rme$ according to $\psi$. 
        Then, the Ring-LWE encryption of a plaintext $\rmm\in R_q$
        is defined by
        \begin{equation*}
            \Enc( \rmm ) := \begin{bmatrix}
                \sk\cdot \rma+\rmm+\rme \\ \rma
            \end{bmatrix}\!\!\mod q  \in R_q^2.
        \end{equation*}
        
        \item ($\mathtt{Decryption}$) $\Dec: R_q^2 \rightarrow R_q$. The decryption of a ciphertext $\cc\in R_q^2$
        is defined by 
        \begin{equation*}
            \Dec(\cc ) := \begin{bmatrix}
                 1 \!\!& -\sk
            \end{bmatrix} \cc \!\! \mod q \in R_q.
        \end{equation*}
         \item ($\mathtt{Homomorphic\ addition}$) $\oplus: R_q^2 \times R_q^2 \rightarrow R_q^2$. The addition of $\cc_1\in R_q^2$ and $\cc_2\in R_q^2$ is defined by 
         \begin{equation*}
             \cc_1 \oplus \cc_2 := \cc_1+\cc_2 \!\!\! \mod q \in R_q^2.
         \end{equation*}
\end{itemize}

\noindent The following proposition is immediate from the definitions of $\Enc$, $\Dec$, and $\oplus$.

\begin{prop1}\label{prop:homoProperties}\upshape
       The following properties hold:
      \begin{enumerate}[leftmargin=*]
        \item For any $\rmm \in R_q$, $\Dec(\Enc(\rmm)) = \rmm + \rme  \!\! \mod q$ for some $\rme\in R_q$ such that $\| \rme \| \le \sigma$.
        \item For any $\cc_1\in R_q^2$ and $\cc_2\in R_q^2$, $\Dec(\cc_1 \oplus \cc_2) = \Dec(\cc_1) + \Dec(\cc_2) \!\! \mod q $.
        \item For any $\rmk \in R_q$ and $\cc\in R_q^2$, $\Dec(\rmk\cdot \cc\!\!\mod q) = \rmk \cdot \Dec(\cc) \!\!\mod q$.\hfill $\square$
    \end{enumerate}
\end{prop1}

The property 2) of Proposition~\ref{prop:homoProperties} is known as the additively homomorphic property of HE.
Along with the property 1), it signifies how the injected errors accumulate under homomorphic additions;
suppose $\rme_1\in R_q$ and $\rme_2\in R_q$ are the errors injected during the encryption of $\rmm_1\in R_q$ and $\rmm_2\in R_q$, respectively. 
It follows that 
\begin{equation*}
    \Dec ( \Enc (\rmm_1) \oplus \Enc(\rmm_2) ) \!=\! (\rmm_1+\rmm_2 ) + (\rme_1+\rme_2 ) \!\!\! \mod q,
\end{equation*}
where the errors are added together.

\begin{rem1}\label{remark 1}\upshape
       In practice, an encrypted message can be recovered without error as follows: 
       For $\rmm\in R_q$ and $\cc\in R_q^2$, let the encryption and decryption algorithms be modified as $ \Enc_\LLL(\rmm):=\Enc( \rmm / \LLL \!\! \mod q )$ and $ \Dec_\LLL(\cc):= \Ceil \LLL \cdot \Dec(\cc)\Flr \!\!\mod q,$  
        where $1/\LLL \in \N$ is some scale factor.
        If $\LLL$ is chosen sufficiently small such that $1/\LLL > 2 \sigma$, then 
    \begin{equation*}
        \Dec_\LLL(\Enc_\LLL(\rmm) )= \Ceil \LLL \cdot \left(\rmm/\LLL +\rme \!\! \mod q \right) \Flr = \rmm,
    \end{equation*}
    as long as $\|\rmm\|<\LLL q/2-1/2$.\hfill $\square$
\end{rem1}

Next, we introduce the Ring-GSW encryption \cite{Chillotti 16}, and describe a method to perform multiplication over encrypted data referred as the ``external product.''
\begin{itemize}[leftmargin=*]
    \item ($\mathtt{Setup}$) Choose $d\in\mathbb{N}$ and a power of two $\nu\in\mathbb{N}$ such that $\nu^{d-1} < q \le \nu^d$.
    \item ($\mathtt{Ring\text{-}GSW \ encryption}$) $\Enc': R_q \rightarrow R_q^{2 \times 2d}$. The Ring-GSW encryption of a plaintext $\rmM\in R_q$
    is defined by
    \begin{equation*}
        \Enc'(\rmM ) := \rmM \cdot \mathcal{G} + \mathcal{L} \!\!\!\mod q \in R_q^{2 \times 2d},
    \end{equation*}
     where $ \mathcal{G} :=\begin{bmatrix}
                     1 & \nu & \cdots & \nu^{d-1}
                 \end{bmatrix} \otimes I_2\in \Z^{2 \times 2d} $ and $\mathcal{L} :=\begin{bmatrix}
                \Enc(0)&\cdots& \Enc(0)
                \end{bmatrix} \in R_q^{2 \times 2d}.$  
\end{itemize}

The external product multiplies a Ring-LWE ciphertext (which belongs to $R_q^2$) with a Ring-GSW ciphertext (which belongs to $R_q^{2\times 2d}$) and returns a Ring-LWE ciphertext.
To define the external product, we begin by introducing the decomposition of a Ring-LWE ciphertext.

\begin{itemize}[leftmargin=*]
    \item ($\mathtt{Ring\text{-}LWE \ ciphertext \ decomposition}$) $\Decomp: R_q^2\rightarrow R_q^{2d}$. 
     For $\cc\in R_q^2$, compute $\bar{\cc}_i\in R_q^2$ for $i=0,1,\ldots,d-1$ such that 
     $\cc = \textstyle\sum_{i=0}^{d-1}\bar{\cc}_i\cdot\nu^i$ and $\| \bar{\cc}_i \| \le \nu/2$.
     The decomposition of $\cc\in R_q^2$ is defined by 
    \begin{equation}\label{eq:decomposition}
        \Decomp(\cc) := [\mathbf{\bar{c}}_{0} ; \cdots  ; \mathbf{\bar{c}}_{d-1}] \in R_q^{2d}.
    \end{equation}
    \item ($\mathtt{External \ product}$)  $\boxdot:R_q^{2 \times 2d} \times R_q^2 \rightarrow R_q^2  $. The external product between $\CC \in R_q^{2 \times 2d}$ and $\cc \in R_q^2$ is defined by 
    \begin{equation*}
        \CC \boxdot \cc := \CC \Decomp(\cc )   \!\! \mod q \in R_q^2.
    \end{equation*}
\end{itemize}
Note that any $\cc \in R_q^2$ can be recovered from the decomposition $\Decomp(\cc )$ as $\cc = \mathcal{G} \Decomp(\cc )$.

The following proposition states the multiplicatively homomorphic property of the external product.

\begin{prop1}\label{prop:multErr}\upshape
       For any $\rmM\in R_q$ and $\cc \in R_q^2$,
       \begin{equation}\label{eq:multHomo}
    \Dec(\Enc'(\rmM)\boxdot\cc
           ) = \rmM \cdot\Dec(\cc)+\Delta  \!\! \mod q
       \end{equation} 
       for some $\Delta\in R_q$ such that $\|\Delta\| \le dN \sigma\nu=:\sigma_{\Mult}$.\hfill $\square$
\end{prop1}

\textit{Proof:} 
It follows from the definitions of $\Enc'$ and $\boxdot$ that
\begin{equation*}
    \begin{split} 
        \Dec(\Enc'(\rmM) \boxdot \cc) 
      &= \begin{bmatrix}
        1 \!&\! -\sk
    \end{bmatrix}  (\rmM \cdot \mathcal{G} + \mathcal{L})  \Decomp ( \cc )  \!\! \mod q \\
      &= \rmM\cdot\Dec(\cc) + \begin{bmatrix}
        1 \!\!& -\sk
    \end{bmatrix}   \mathcal{L}  \Decomp ( \cc )  \!\! \mod q.
    \end{split}
\end{equation*}  
\noindent 
Here, $\begin{bmatrix}
    1  \!\!& -\sk
\end{bmatrix}   \mathcal{L}$ is a $2d$-dimensional row vector bounded as $\| \begin{bmatrix}
    1  \!&\! -\sk
\end{bmatrix}   \mathcal{L} \| \le 2d\sigma$. 
Since $\Decomp ( \cc )\in R_q^{2d}$ is also bounded as $\|\Decomp(\cc) \|\le\nu/2$, it follows from \eqref{eq:defAddMult} that $ \| \begin{bmatrix}
        1 \!\!& -\sk
    \end{bmatrix}   \mathcal{L}  \Decomp ( \cc ) \|\le dN \sigma\nu.$ 
This concludes the proof.
\hfill$\blacksquare$

In general, $\cc$ in \eqref{eq:multHomo} is not necessarily a ``freshly encrypted'' ciphertext, but rather could have already gone through recursive homomorphic operations, thus carrying some accumulated error.
Nevertheless, Proposition~\ref{prop:multErr} guarantees that the newborn error under the external product, which is $\Delta$ in \eqref{eq:multHomo}, is bounded by a constant regardless of the error accumulated in $\cc$.

For simplicity, we abuse notation and define $\Enc(\cdot)$ and $\Enc'(\cdot)$ component-wisely with respect to vectors or matrices of plaintexts.
The operation $\oplus$ and $\Dec(\cdot)$ is also applied component-wisely to vectors or matrices of ciphertexts.
For $\rmK=\{\rmK_{i,j} \}\in R_q^{h\times l}$ and $\cc=[\cc_1;\cdots;\cc_l ] \in R_q^{2l}$, the external product between $\Enc'(\rmK)\in R_q^{2h \times 2ld}$ and $\cc$ is defined by 
\begin{equation*}
    \begin{split}
        &\Enc'(\rmK)\boxdot \cc \\
        &\!= \!\left[ \textstyle\sum_{j=1}^l \Enc' (\rmK_{1,j}) \boxdot   \cc_j;\!\cdots\!; \textstyle\sum_{j=1}^l  \Enc' (\rmK_{h,j})\boxdot \cc_j \right]\!\!\in\!\! R_q^{2h}\!,
    \end{split}
\end{equation*}
where the summation is taken with respect to the operation $\oplus$.
Then, it follows from Proposition~\ref{prop:multErr} that 
\begin{align}\label{eq:multHomoMat}       
        &\Dec(\Enc'(\rmK)\boxdot \cc ) \nonumber \\
        &\!\!=\!\left[\textstyle\sum_{j=1}^l \! \rmK_{1,j}  \!\cdot\!  \Dec(\cc_j) ;\! \cdots \!;\! \textstyle\sum_{j=1}^l \!  \rmK_{h,j} \!\cdot\! \Dec(\cc_j)  \right]  \!+ \!\Delta \!\!\!\! \mod q \nonumber  \\
        &\!\!= \rmK\cdot \Dec(\cc)+\Delta \!\!\!\ \mod q \in R_q^h
\end{align}
for some $\Delta\in R_q^h$ such that $\| \Delta \| \le l\cdot\sigma_{\Mult}$.

\subsection{Problem formulation}

Consider a discrete-time plant written by
\begin{equation} \label{eq:plant}
        \begin{split}
        x_p(t+1)&=Ax_p(t)+Bu(t), \quad x_p(0)=x_{p}^{\ini},\\
        y(t)&=Cx_p(t),
        \end{split}
\end{equation}  
where $x_p(t)\in\mathbb{R}^{n_p}$ is the state with the initial value $x_p^{\mathsf{ini}}\in\mathbb{R}^{n_p}$, $u(t)\in\mathbb{R}^{m}$ is the input, and $y(t)\in\mathbb{R}^{p}$ is the output.

Suppose that a controller stabilizing the plant \eqref{eq:plant} has been designed as 
\begin{equation}\label{eq:nominalController}
        \begin{split}
        x(t+1)&=Fx(t)+Gy(t), \quad x(0)=x^{\ini},\\
        u(t)&=Hx(t),
        \end{split}
\end{equation}  
where $x(t)\in\R^n$ is the state with the initial value $x^{\ini}\in \Q^n$, and the control parameters are given as
\begin{equation*}
    F\in\Z^{n\times n},\quad G\in\Q^{n\times p},\quad \text{and}\quad H\in\Q^{m\times n}.
\end{equation*}
It has been studied in \cite{Cheon 18} that the state matrix, such as $F$ of \eqref{eq:nominalController}, needs to be an integer matrix, in order to implement a dynamic system over encrypted data.
Subsequently, there have been studies \cite{Kim 21, Tavazoei 23, Lee 23}  on converting a given controller to have an integer state matrix, while preserving its control performance\footnote{
The conversion is always possible through approximation \cite{Kim 21}. Other results \cite{Tavazoei 23, Lee 23} avoid such approximation under some systematic assumptions.}.
Throughout this paper, we assume that such conversion has been conducted a priori.
Moreover, the elements of
$\{x^{\ini},G,H\}$
being rational numbers is fairly reasonable because real numbers can be approximated to rational numbers with arbitrary precision, as $\Q$ is dense in $\R$.    

The objective is to implement \eqref{eq:nominalController} as an encrypted controller, which satisfies the followings:
\begin{itemize}[leftmargin=*]
    \item The encrypted controller performs an unlimited number of recursive homomorphic multiplications, on the Ring-LWE based cryptosystem described in Section~\ref{subsec:rlwe}.
    \item It operates for an infinite time horizon, without bootstrapping or any other methods to refresh the (encrypted) state.
    \item The performance of the proposed controller is equivalent to that of \eqref{eq:nominalController} in the following sense: Let $u(t)$ be the control input generated by the encrypted controller and $u^{\nom}(t)$ be the input from the pre-designed controller \eqref{eq:nominalController}. Given $\epsilon>0$,
 \begin{equation}\label{eq:problem}
  \left\| u(t)- u^\nom(t)\right\|\le\epsilon
 \end{equation}
    holds for all $t\in\Z_{\ge 0}$.
\end{itemize}

\section{Encrypted Controller}\label{Sec:Naive}
In this section, we present a method to operate the controller \eqref{eq:nominalController} using the Ring-LWE based scheme.
It is shown that the effect of error growth is suppressed by the closed-loop stability, enabling the proposed encrypted controller to execute an unlimited number of recursive homomorphic multiplications without the use of bootstrapping.
Additionally, we provide an explicit guideline for selecting parameters to achieve \eqref{eq:problem}.

\subsection{Encrypted controller design}\label{Sec:NaiveEnccontroller}

To implement the operations of \eqref{eq:nominalController} over encrypted data, we utilize the homomorphic properties stated in Propositions~\ref{prop:homoProperties} and \ref{prop:multErr}.
Since the plaintext space of the Ring-LWE based cryptosystem is a polynomial ring, each component of the matrices and vectors in \eqref{eq:nominalController} is regarded as a constant polynomial. 

During the offline procedure, the control parameters---$F$, $G$, and $H$ of \eqref{eq:nominalController}---and the initial state $x^{\ini}$ of the controller are quantized and encrypted as follows; 
let $\rrr >0$ be the quantization step size for the plant output at the sensor.
As the elements of $G$, $H$, and $x^{\ini}$ belong to $\Q$, a scale factor $1/\sss \in \mathbb{N}$ can be chosen such that
\begin{align}\label{eq:quantizedParams}
    G/\sss&\in\mathbb{Z}^{n \times p}, &
        H/\sss&\in\mathbb{Z}^{m \times n}, &
        x^{\ini}/(\rrr\sss)&\in\mathbb{Z}^{n}.
\end{align}
We also introduce a parameter $1/\LLL\in\mathbb{N}$, in order to deal with the effect of errors as illustrated in Remark~\ref{remark 1}.
Then, the control parameters and the initial state are encrypted as
\begin{align}\label{eq:encFGH}
          \FF&:=\Enc'(F \!\!\!\mod q), &
    \GG&:=\Enc' (G/\sss \!\!\!\mod q) ,  \\
    \HH&:=\Enc'(H/\sss \!\!\!\mod q) , &
    \xx^{\ini}&:=\Enc (x^{\ini} / (\rrr\sss\LLL) \!\!\!\mod q) . \nonumber
\end{align}
The rationale behind this is that the external product between a Ring-GSW ciphertext and a Ring-LWE ciphertext returns a Ring-LWE ciphertext,
as the product between a matrix and a vector returns a vector.

The encrypted control system operates as follows:
At each time step $t\in\Z_{\ge 0}$ of the online procedure, the sensor quantizes the plant output $y(t)$ as
\begin{equation}\label{eq:ypreprocess}
     \bar{y}(t):=\left\lceil \frac{y(t)}{\rrr}\right\rfloor\in \mathbb{Z}^p,
\end{equation}
and then encrypts it as
\begin{equation}\label{eq:ency}
    \yy(t):=\Enc(\bar{y}(t) / \LLL \!\!\!\mod q)\in R_q^{2p},
\end{equation}
which becomes the input of the encrypted controller.
The proposed controller is described as
\begin{subequations}\label{eq:encController}
\begin{equation}\label{eq:encControllerDyn}
\begin{aligned}
    \xx(t+1)&=\left(\FF \boxdot\xx(t)\right)\oplus\left(\GG \boxdot\yy(t)\right), \quad \xx(0)=\xx^{\ini}, \\
        \uu(t)&=\HH\boxdot \xx(t), 
\end{aligned}
\end{equation}
where $\xx(t)\in R^{2n}_{q}$ is the state and $\uu(t)\in R^{2m}_{q}$ is the output.
The output $\uu(t)$ is transmitted to the actuator, decrypted, and then scaled down to obtain the plant input $u(t)$, as 
\begin{equation}\label{eq:encControllerOutput}
    u(t) = \rrr\sss^2\LLL\cdot \Dec(\uu(t) )\big|_{X=0}.
\end{equation}
\end{subequations}



\subsection{Performance analysis}

Here, we analyze the performance of the proposed encrypted controller \eqref{eq:encController}, and show that \eqref{eq:problem} holds with an appropriate choice of the parameters $\{q,\rrr,\sss,\LLL\}$.
To this end, the growth of the error coefficients, which are sampled as in \eqref{eq:RLWEerr} and injected during the encryption of ciphertexts in \eqref{eq:encController}, should be examined.
Specifically, it is shown that 
i) the error coefficients in the non-constant terms do not affect the control performance in the proposed design; 
ii) the effect of the error coefficients in the constant terms remain bounded by the closed-loop stability of \eqref{eq:plant} with \eqref{eq:nominalController}, and therefore, \eqref{eq:problem} can be achieved.

For analysis, we first define
\begin{equation}\label{eq:decStateOutput}
\begin{aligned}
    \rmu(t)&:=\rrr\sss^2\LLL\cdot \Dec(\uu(t))=\textstyle\sum_{i=0}^{N-1}\rmu_i(t)X^i,\\
    \rmx(t)&:=\rrr\sss\LLL\cdot\Dec(\xx(t))=\textstyle\sum_{i=0}^{N-1}\rmx_i(t)X^i,
\end{aligned}
\end{equation}
where $\rmu_i(t)\in\mathbb{R}^m$ and $\rmx_i(t)\in\mathbb{R}^n$, for $i=0,1,\dots,N-1$, are the vectors consisting of the $i$-th coefficients.
Note that $\rmu_0(t)\equiv u(t)$
of \eqref{eq:encControllerOutput} and we interchangeably use both notations in this section.

Next, we define some ``lumped errors'' that are generated by quantization and encryption algorithms, written by
\begin{align}\label{eq:perturbationsDef}
        \rme^x(t) &:= \rmx(t+1) - \left(F\rmx(t) + Gy(t)\right), &
         \rme^{\ini} &:= \rmx(0)-x^\ini, \nonumber \\
        \rme^u(t) &:= \rmu(t)-H\rmx(t).
\end{align}
Likewise, the vectors consisting of the $i$-th coefficients of $\rme^x(t)$, $\rme^u(t)$, and $\rme^{\ini}$ are denoted by $\rme_i^x(t)\in\mathbb{R}^n$, $\rme_i^u(t)\in\mathbb{R}^m$, and $\rme_i^{\ini}\in\mathbb{R}^n$, respectively, for $i=0,1,\dots,N-1$.

Then, the dynamics of each coefficient vector $\rmx_i(t)$ can be derived from \eqref{eq:perturbationsDef}, as
\begin{align}\label{eq:constantController}
        \rmx_0(t+1)&=F\rmx_0(t)+Gy(t)+\rme_0^x(t), & \rmx_0(0)&=x^{\ini}+\rme_0^{\ini},\nonumber \\
        \rmu_0(t)&=H\rmx_0(t)+\rme_0^u(t),
\end{align}
and for $i=1,\dots,N-1$,
\begin{align*}
    \rmx_i(t+1)&=F\rmx_i(t)+\rme_i^x(t), \quad \rmx_i(0)=\rme_i^{\ini},\\
        \rmu_i(t)&=H\rmx_i(t)+\rme_i^u(t).
\end{align*}
The reason why the dynamics of $\rmx_0(t)$ differs from those of other $\rmx_i(t)$'s is that the components of $y(t)$ and $x^{\ini}$ are scalars, which only affect the constant terms of $\rmx(t)$.

Since the output $u(t)\equiv\rmu_0(t)$ of \eqref{eq:encControllerOutput} is solely determined from \eqref{eq:constantController}, the closed-loop system of \eqref{eq:plant} and \eqref{eq:encController} can be divided into two parts:
\begin{equation}\label{eq:decoupleHigh}
    \begin{split}
         \begin{bmatrix}
                \rmx_1(t+1) \\
                \xvdots \\
                \rmx_{N-1}(t+1)
            \end{bmatrix}
            \!&=\! I_{N-1} \!\otimes\! F
            \begin{bmatrix}
                \rmx_1(t) \\
                \xvdots \\
                \rmx_{N-1}(t)
            \end{bmatrix} \!+\! \begin{bmatrix} 
                            \rme^x_1(t)  \\ 
                            \xvdots \\
                            \rme^x_{N-1}(t) 
            \end{bmatrix}, \\
            \rmx_i(0) &= \rme_i^\ini, \quad i=1,\ldots,N-1,
    \end{split}
\end{equation} 
where the coefficients in the non-constant terms accumulate under the recursive update of $\rmx_i(t)$'s, and
\begin{align}\label{eq:closedLoop}
            \!\!\!\!\begin{bmatrix} 
                            x_p(t+1)  \\ 
                           \rmx_0(t+1) \\
            \end{bmatrix} &=
            \underbrace{\begin{bmatrix} 
                            A  & BH  \\ 
                            GC &F 
            \end{bmatrix}}_{=:\bar{A}} 
             \begin{bmatrix} 
                             x_p(t) \\ 
                            \rmx_0(t) \\
            \end{bmatrix} 
             +\begin{bmatrix}
                 B & 0 \\
                 0 & I_n
             \end{bmatrix}\begin{bmatrix} 
                            \rme^u_0(t)  \\ 
                            \rme^x_0(t) 
            \end{bmatrix}, \nonumber
           \\
            \begin{bmatrix}
                x_p(0) \\
                \rmx_0(0)
            \end{bmatrix} &= \begin{bmatrix}
                x_{p}^{\ini} \\
                x^{\ini} + \rme_0^{\ini}
            \end{bmatrix},
\end{align}
which is only relevant to the coefficients in the constant terms.
This shows that the performance of the encrypted controller \eqref{eq:encController} is equivalent to that of the controller \eqref{eq:constantController} over $\mathbb{R}$.

The following lemma states that $\rme_0^x(t)$, $\rme_0^u(t)$, and $\rme_0^{\ini}$ of \eqref{eq:closedLoop}, which can be thought of as perturbations, remain bounded for all $t\in \Z_{\ge 0}$ if the modulus $q$ is chosen sufficiently large. 
To state the lemma, some technical definitions are made.
Since the closed-loop state matrix $\bar{A}$ of \eqref{eq:closedLoop} is Schur stable by the assumption on \eqref{eq:nominalController}, there exists $ M>0$ and $0\le \lambda<1$ such that $\| \bar{A}^k \| \le M \lambda^k$ for all $k\ge0$ \cite[Theorem 3.5]{Dowler 13}.
We define the function $\eta:\mathbb{R}\times\mathbb{R}\times\mathbb{R}\rightarrow \mathbb{R}$ by 
\begin{equation}\label{eq:etaDef}
    \eta(\alpha, \beta, \gamma) :=  M\left( \left\| 
    \begin{bmatrix}
        x_{p}^{\ini} \\
            x^{\ini}
    \end{bmatrix}
    \right\|+\gamma + \frac{ \| B \| \beta + \alpha}{1-\lambda}  \right).
\end{equation}

\begin{lem1}\label{lemma1}\upshape
   Given the parameters $\rrr>0$, $1/\sss\in\N$, and $1/\LLL\in\N$, the perturbations of \eqref{eq:closedLoop} remain bounded by 
    \begin{align}\label{eq:lem1Show}
        \| \rme_0^x(t) \| \!&\le\! \alpha(\rrr,\sss,\LLL):= \rrr\sss\LLL(n+p) \sigma_{\Mult} \!+\!\rrr\frac{\| G \|}{2} \!+\! \rrr\LLL\| G \| \sigma, \nonumber\\
        \| \rme_0^u(t)  \| \! &\le \! \beta(\rrr,\sss,\LLL):=\rrr\sss^2\LLL n \sigma_{\Mult},  \\ 
        \| \rme_0^{\ini}  \| \!&\le\! \gamma(\rrr,\sss,\LLL):=\rrr\sss\LLL\sigma, \nonumber
    \end{align}
    for all $t\in \Z_{\ge 0}$ if 
    \begin{equation}\label{eq:lem1qbound}
         q > 2\max \left\{\frac{\bar{\eta}(\rrr,\sss,\LLL)}{\rrr\sss\LLL}, \frac{\| H \| \bar{\eta}(\rrr,\sss,\LLL)+  \beta(\rrr,\sss,\LLL)}{\rrr\sss^2\LLL}   \right\},
    \end{equation}
    where $\bar{\eta}(\rrr,\sss,\LLL) := \eta(\alpha(\rrr,\sss,\LLL),\beta(\rrr,\sss,\LLL),\gamma(\rrr,\sss,\LLL))$.\hfill $\square$
\end{lem1}

\textit{Proof:}
    See Appendix~\ref{appendix Lemma 1}. 
\hfill $\blacksquare$

Lemma~\ref{lemma1} implies that the perturbations $\rme_0^x(t)$, $\rme_0^u(t)$, and $\rme_0^{\ini}$ can be arbitrarily bounded by decreasing $\rrr$, $\sss$, and $\LLL$, because their upper bound functions $\alpha(\rrr,\sss,\LLL)$, $\beta(\rrr,\sss,\LLL)$, and $\gamma(\rrr,\sss,\LLL)$ vanish at the origin.
Making use of this result, the following theorem states that the performance error in \eqref{eq:problem} can be made arbitrarily small with the choice of the parameters $\{q,\rrr,\sss,\LLL\}$.

\begin{thm1}\label{theorem1}\upshape
    For given $\epsilon>0$, if the parameters $\rrr>0$, $1/\sss\in\N$, and $1/\LLL\in\N$ satisfy 
    \begin{align}\label{eq:thm1Bound}
        \!\!\!\!\beta(\rrr,\sss,\LLL )\le \frac{\epsilon}{2} \ \ \ \mbox{and} \ \ \ 
            \bar{\eta}(\rrr,\sss,\LLL)  \le  M\left\|  \begin{bmatrix}
        x_{p}^{\ini} \\
            x^{\ini}
    \end{bmatrix} \right\| + \frac{\epsilon}{2\left\|H \right\|}, 
    \end{align}
     and the modulus $q$ satisfies \eqref{eq:lem1qbound}, then the encrypted controller \eqref{eq:encController} guarantees that \eqref{eq:problem} holds
     for all $t \in \Z_{\ge 0} $. \hfill $\square$
\end{thm1}

\textit{Proof:}
    See Appendix~\ref{appendix Theorem 1}. 
\hfill $\blacksquare$

Note that \eqref{eq:thm1Bound} can be satisfied by choosing a sufficiently small $\rrr$, $\sss$, and $\LLL$.
We suggest that $\rrr$ be chosen prior to $\sss$ and $\LLL$ because decreasing $\rrr$ may require higher sensor resolution in practice, which could be costly. 
After choosing $\rrr$, $\sss$, and $\LLL$, the modulus $q$ can be chosen to satisfy \eqref{eq:lem1qbound}.

\begin{rem1}
    The proposed Ring-LWE based encrypted controller \eqref{eq:encController} exhibits lower time complexity compared to the LWE based encrypted controller presented in \cite{Kim 23}. 
    At each time step, both controllers perform $(n^2+np+nm)$-external products, however, the time complexities of the external product \cite{Chillotti 16} used in \eqref{eq:encController} and the external product \cite{Gentry 13} used in \cite{Kim 23} are given by $\calO(dN\log(N))$ and $\calO(dN^2)$, respectively. \hfill $\square$
\end{rem1}

\section{Encrypted Controller with Packing}\label{Sec:Hoisted}

In this section, we propose a novel packing algorithm that encodes a vector into a polynomial, which significantly reduces the number of operations required for a matrix-vector multiplication.
Then, we present an encrypted controller that integrates the packing algorithm, offering improved computational and memory efficiency compared to the encrypted controller \eqref{eq:encController}.
We again analyze and show that the effect of error growth can be arbitrarily bounded by means of the closed-loop stability.   

For notational convenience, we often omit the modulo operation $\!\!\!\mod q$ in this section.

\subsection{Packing  and unpacking algorithms}

Let us first define
\begin{equation}\label{eq:deftau}
    \tau := 2^{\left\lceil \log_2(\max\{n,m,p\}) \right\rceil } \in \N,
\end{equation}
which is the smallest power of two such that $\tau \ge \max\{n,m,p \}$.
It is assumed that $N\ge \tau$, which can be easily satisfied by increasing $N$.

\subsubsection{Packing}
A packing algorithm that converts a vector of integers in $\Z_q$ into a polynomial in $R_q$ is presented. 

\begin{itemize}[leftmargin=*]
    \item $\Pack_k: \Z_q^k \to R_q$. For $k\in\N$ such that $k\le \tau$, the packing of a vector $a=\left[a_0;a_1;\cdots;a_{k-1} \right] \in\mathbb{Z}_q^k$ is defined by 
    \begin{equation*}
        \Pack_k(a ) := a_{0} + a_{1}X^{N/\tau} +  \cdots + a_{k-1}X^{(k-1)N/\tau} \in R_q.
    \end{equation*}
\end{itemize}

\noindent The packing algorithm $\Pack_k$ assigns each element of a vector as the coefficients of the terms $X^0, X^{N/\tau}$, \ldots, $X^{(\tau-1)N/\tau}$ in sequence.
These terms will be referred to as \textit{packing slots} in this paper.

A matrix-vector multiplication can be expressed with polynomial additions and multiplications using the packing algorithm.
For $A\in \Z_q^{h\times l}$ and  $b\in\Z_q^l$ such that $h\le \tau$ and $l\le \tau$, consider the matrix-vector multiplication
\begin{equation}\label{eq:matvecAB}
    A b =\textstyle\sum_{i=0}^{l-1} A_i\cdot b_i\in\Z_q^h,
\end{equation}
where $A_i$ denotes the $(i\!+\!1)$-th column of $A$ and $b_i$ denotes the $(i\!+\!1)$-th element of $b$. 
Then, it holds that
\begin{equation}\label{eq:packAb}
    \Pack_h(Ab ) = \textstyle\sum_{i=0}^{l-1} \Pack_h(A_i)\cdot b_i.
\end{equation}
In fact, the right-hand side can be evaluated homomorphically with $l$ external products and $l-1$ homomorphic additions, as 
\begin{equation}\label{eq:packHomo}
    \textstyle\sum_{i=0}^{l-1} \Enc'(\Pack_h(A_i))\boxdot \Enc(b_i).
\end{equation}

\subsubsection{Unpacking plaintexts (polynomials)}
An unpacking algorithm that converts a polynomial in $R_q$ into a vector of integers in $\Z_q$ is presented.
This can be employed to restore the vector $Ab$ from $\Pack_h(Ab)$ in \eqref{eq:packAb}.
\begin{itemize}[leftmargin=*]
    \item $\UnpackPt_k: R_q \rightarrow \Z_q^k $. For $k\in\N$ such that $k\le \tau$, the unpacking of a polynomial $\rmm = \textstyle\sum_{i=0}^{N-1}\rmm_iX^i\in R_q$
    is defined by 
    \begin{equation*}
        \UnpackPt_k(\rmm) := \left[\rmm_0;\rmm_{N/\tau};\cdots;\rmm_{(k-1)N/\tau} \right]\in\mathbb{Z}_q^k.
    \end{equation*}
\end{itemize}

\noindent The unpacking algorithm $\UnpackPt_k$ constructs a $k$-dimensional vector from the first $k$ packing slots of a given polynomial. 
Hence, applying $\UnpackPt_h$ to \eqref{eq:packAb} yields
\begin{equation}\label{eq:AbForm}
        Ab = \UnpackPt_h ( \textstyle\sum_{i=0}^{l-1} \Pack_h(A_i)\cdot b_i).
\end{equation}
This implies that $Ab$ can be evaluated homomorphically using \eqref{eq:packHomo} if $\UnpackPt_h$ can be evaluated homomorphically.
By this motivation, we delve deeper into the unpacking algorithm.

\begin{algorithm}[t]
\caption{Unpacking plaintexts}
    \begin{algorithmic}[1]\label{alg:UnpackPt}
        \REQUIRE 
         $\rmm(X)\in R_q$ of the form \eqref{eq:packedPoly}
        
        \renewcommand{\algorithmicrequire}{\textbf{Procedure}}
        \REQUIRE $\UnpackPt_k(\rmm(X))$
        \STATE $\rmm_0 \gets \rmm$      
        \FOR{$(\zeta=\tau;\zeta>1; \zeta = \zeta/2)$} 
            \FOR{$(\omega=0; \omega<\tau; \omega=\omega+\zeta)$}
                \STATE $\rmm_\omega \gets (q+1)/2\cdot \rmm_\omega $
                \STATE $\mathrm{tmp}_\omega \gets \rmm_\omega(X^{\zeta+1})$
                \STATE $\rmm_{\omega+\zeta/2} \gets (\rmm_\omega-\mathrm{tmp}_\omega)\cdot X^{-N/\zeta} $
                \STATE $\rmm_\omega \gets \rmm_\omega+\mathrm{tmp}_\omega$
            \ENDFOR
        \ENDFOR

        \FOR{$(\idx = 0; \idx < \tau; \idx = \idx + 1)$}
            \STATE $\BRidx \leftarrow \Bitreverse(\idx)$
            \IF{$(\idx < \BRidx)$}
                \STATE $\rmm_\idx, \rmm_\BRidx \leftarrow \rmm_\BRidx, \rmm_\idx$
            \ENDIF
        \ENDFOR
        \RETURN $\left[\rmm_0;\dots;\rmm_{k-1} \right]\in \Z_q^k$ 
    \end{algorithmic}
\end{algorithm}

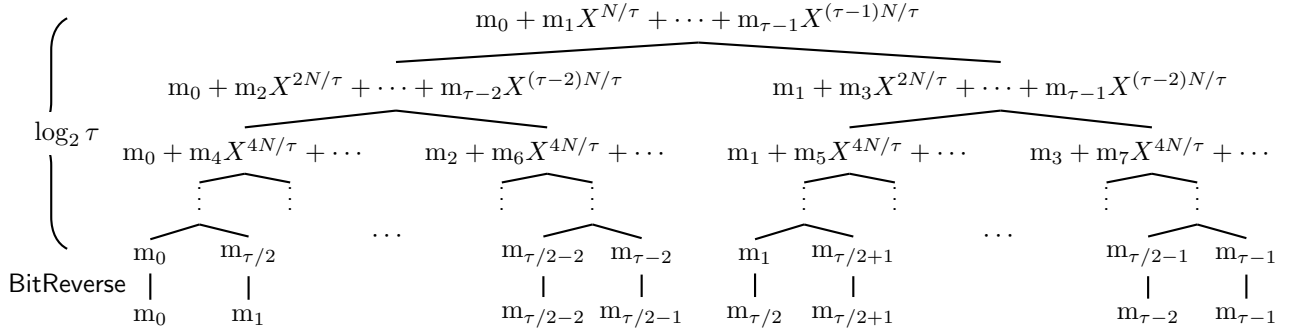
\begin{figure*}
    \begin{center}
        \begin{tikzpicture}[
            level 1/.style={sibling distance=80mm, level distance=9mm},
            level 2/.style={sibling distance=40mm, level distance=8.5mm},
            level 3/.style={sibling distance=12mm, level distance=7mm},
            level 4/.style={sibling distance=13mm, level distance=7mm},
            level 5/.style={sibling distance=13mm, level distance=8mm},
            text=black,
            >=latex,
            font=\sffamily
            ]
    
         \node (z){$\rmm_0+\rmm_1X^{N/\tau}+\cdots + \rmm_{\tau-1}X^{(\tau-1)N/\tau}$} 
             child {node (a) {$\rmm_0+\rmm_2X^{2N/\tau}+\cdots + \rmm_{\tau-2}X^{(\tau-2)N/\tau}$}
                child {node  (b) {$\rmm_0+\rmm_4X^{4N/\tau}+\cdots $}
                    child {node  (b1) {\scriptsize$\xvdots$}
                        child {node  (b11) {$\rmm_0$} child {node  (b111) {$\rmm_0$}} }
                        child {node  (b12) {$\rmm_{\tau/2}$} child {node  (b121) {$\rmm_1$}} }   
                    }
                    child {node  (b2) {\scriptsize$\xvdots$}}
                }
                child {node (g) {$\rmm_2+\rmm_6X^{4N/\tau}+\cdots $}
                    child {node  (g1) {\scriptsize$\xvdots$}
                    }
                    child {node  (g2) {\scriptsize$\xvdots$}
                        child {node  (g21) {$\rmm_{\tau/2-2}$}  child {node  (g211) {$\rmm_{\tau/2-2}$}} 
                        }
                        child {node  (g22) {$\rmm_{\tau-2}$}
                        child {node  (g221) {$\rmm_{\tau/2-1}$}}
                        }  
                    }
                }
            }
            child {node (d) {$\rmm_1+\rmm_3X^{2N/\tau}+\cdots + \rmm_{\tau-1}X^{(\tau-2)N/\tau}$}
              child {node  (e) {$\rmm_1+\rmm_5X^{4N/\tau}+\cdots $}
                child {node  (e1) {\scriptsize$\xvdots$}
                    child {node(e11) {$\rmm_1$} child {node  (e111) {$\rmm_{\tau/2}$}}
                    }
                    child {node(e12) {$\rmm_{\tau/2+1}$} child {node  (e121) {$\rmm_{\tau/2+1}$}}
                    }  
                }
                child {node  (e2) {\scriptsize$\xvdots$}}
                }
              child {node (f) {$\rmm_3+\rmm_7X^{4N/\tau}+\cdots $}
                child {node  (f1) {\scriptsize$\xvdots$}
                }
                child {node  (f2) {\scriptsize$\xvdots$}
                    child {node(f21) {$\rmm_{\tau/2-1}$} child {node  (f211) {$\rmm_{\tau-2}$}}
                    }
                    child {node(f22) {$\rmm_{\tau-1}$} child {node  (f221) {$\rmm_{\tau-1}$}}
                    } 
                }
            }
        };

        \path (b12.north east) -- (g21.north west) node [midway] {$\cdots$};
        \path (e12.north east) -- (f21.north west) node [midway] {$\cdots$};
        
        \coordinate (cd1) at ($(b11)+(-1.1,0.1)$);
        \node[below=0.2 of cd1]{$\Bitreverse$};
        \draw[black,thick,decorate,decoration={brace,amplitude=12pt}] 
            (cd1) -- (cd1|-z.west) node [midway, fill=white] {$\log_2\tau$};
        
        \end{tikzpicture}
    \end{center}
    \caption{Illustration of $\UnpackPt_k(\rmm)$ for $\rmm(X)$ given in \eqref{eq:packedPoly}.} 
    \label{fig:unpack}
\end{figure*}

Consider a polynomial $\rmm(X)\in R_q$ written by
\begin{equation}\label{eq:packedPoly}
    \rmm(X) := \rmm_0 + \rmm_1X^{N/\tau} + \cdots + \rmm_{\tau-1} X^{(\tau-1)N/\tau}.
\end{equation}
Since $X^N$ is regarded as $-1$ in $R_q$, it follows that 
\begin{equation*}\label{automorphism of m}
    \rmm(X^{\tau+1}) =   \rmm_0 - \rmm_1X^{N/\tau} + \cdots - \rmm_{\tau-1} X^{(\tau-1)N/\tau}.
\end{equation*}
Therefore, by adding and subtracting $\rmm(X^{\tau+1})$ from $\rmm(X)$, we can separate the terms, as 
\begin{equation*}
    \begin{split}
       \rmm_E(X) :=&  \frac{q+1}{2}\cdot\rmm(X) +\frac{q+1}{2}\cdot \rmm(X^{\tau+1})  \\
       =& \rmm_0 + \rmm_2X^{2N/\tau} + \cdots + \rmm_{\tau-2} X^{(\tau-2)N/\tau} ,\\
       \rmm_O(X) :=& \left(\frac{q+1}{2}\cdot\rmm(X) - \frac{q+1}{2}\cdot\rmm(X^{\tau+1}) \right)\cdot  X^{-N/\tau}  \\
       =& \rmm_1 + \rmm_3X^{2N/\tau} + \cdots + \rmm_{\tau-1} X^{(\tau-2)N/\tau}.\\
    \end{split}
\end{equation*}
The scalar $(q+1)/2$, which is the multiplicative inverse\footnote{
In $\mathbb{Z}_q \subset R_q$, $q+1=1$. Thus, $(q+1)/2$ is the multiplicative inverse of two because $2\cdot(q+1)/2=q+1=1$.
}
of two in the set $\Z_q$, is pre-multiplied to halve the doubled coefficients. 
The monomial $X^{-N/\tau}$ is multiplied in obtaining $\rmm_O(X)$ to simply match the forms of $\rmm_E(X)$ and $\rmm_O(X)$. 
Each term of $\rmm(X)$ can be extracted by repeating this process, thus obtaining $\UnpackPt_k(\rmm(X))$. 
This is described in Algorithm~\ref{alg:UnpackPt} and also depicted in Fig.~\ref{fig:unpack}.

The operation $\Bitreverse:\Z_{\ge 0}\rightarrow \Z_{\ge 0}$ in Algorithm~\ref{alg:UnpackPt} reverses the ($\log_2 \tau$)-digit binary expression of the input. 
For example, if $\tau=8$ then $\Bitreverse(\tau/2)=\Bitreverse(4)=1$ because the $3$-digit binary expression of $4$ is $[100]_2$, and its reverse $[001]_2$ is $1$ in the decimal expression.
This is necessary to rearrange the indices of the extracted coefficients in sequential order.

\subsubsection{Unpacking ciphertexts}
We now present a method to evaluate $\UnpackPt_k$ homomorphically over encrypted data. 
To this end, we introduce the following definitions.

\begin{itemize}[leftmargin=*]
    \item ($\mathtt{Polynomial \ automorphism}$) $\Psi_\theta: R_q \rightarrow R_q$. For odd $\theta\in\N$, the automorphism of a polynomial $\rmm(X)\in R_q$ is defined by $\Psi_\theta(\rmm(X)) := \rmm(X^\theta) \in R_q.$
    \item ($\mathtt{Automorphism \ key}$) For odd $\theta\in\N$, the automorphism key $\ak_\theta\in R_q^{2\times 2d}$ for the automorphism $\Psi_\theta$ under the secret key $\sk$ is defined by $\ak_\theta := \Enc' (\Psi_\theta(\sk))$.
    \item ($\mathtt{Ring\text{-}LWE \ ciphertext \ automorphism}$) $\Phi_\theta:R_q^2 \times R_q^{2 \times 2d} \rightarrow R_q^2 $. For odd $\theta\in\N$, the automorphism of $\cc=[\rmb;\rma]\in R_q^2$ is defined by 
    \begin{equation*}
        \Phi_\theta(\cc, \ak_\theta ) \!:=\! [\Psi_\theta(\rmb) ; 0]- \ak_\theta \boxdot [\Psi_\theta(\rma) ; 0]\!\!\!\!\mod q \in R_q^2.
    \end{equation*}
\end{itemize}

\begin{algorithm}[t]
\caption{Unpacking ciphertexts}
    \begin{algorithmic}[1]\label{alg:UnpackCt}
        \REQUIRE $\cc\in R_q^2$

       \renewcommand{\algorithmicrequire}{\textbf{Procedure}}
        \REQUIRE $\UnpackCt_k(\cc)$
        \STATE Prepare $\mathrm{ak}_\theta$ for $\theta \in \{2^\xi+1 \mid 2 \le 2^\xi \le  \tau, \ \xi\in\N \}$
        \STATE $\cc_0 \gets \cc$
        
        \FOR{$(\zeta=\tau;\zeta>1; \zeta = \zeta/2)$} 
            \FOR{$(\omega=0; \omega<\tau; \omega=\omega+\zeta)$}
                \STATE $\cc_\omega \gets (q+1)/2\cdot \cc_\omega $
                \STATE $\mathbf{tmp}_\omega \gets \Phi_{\zeta+1}(\cc_\omega,\mathrm{ak}_{\zeta+1})$
                \STATE $\cc_{\omega+\zeta/2} \gets (\cc_\omega \oplus (-\mathbf{tmp}_\omega))\cdot X^{-N/\zeta} $
                \STATE $\cc_\omega \gets \cc_\omega\oplus \mathbf{tmp}_\omega$\label{line8}
            \ENDFOR
        \ENDFOR 

        \FOR{$(\idx = 0; \idx < \tau; \idx = \idx + 1)$}
            \STATE $\BRidx \leftarrow \Bitreverse(\idx)$
            \IF{$(\idx < \BRidx)$}
                \STATE $\cc_\idx, \cc_\BRidx \leftarrow \cc_\BRidx, \cc_\idx$
            \ENDIF
        \ENDFOR
        \RETURN $\left[\cc_0;\dots;\cc_{k-1} \right]\in R_q^{2k}$ 
    \end{algorithmic}
\end{algorithm}

\noindent The following proposition demonstrates the homomorphic property of the automorphism $\Phi_\theta$.

\begin{prop1}\label{prop:autom}\upshape
    For any $\cc=[\rmb;\rma]\in R_q^2$ and odd $\theta\in\N$, 
    \begin{equation}\label{lemma 2 expression}
        \Dec(\Phi_\theta (\cc,\ak_\theta ) ) = \Psi_\theta(\Dec(\cc) ) + \Delta \!\!\mod q
    \end{equation}
    for some $\Delta\in R_q$ such that $\|\Delta \| \le dN \sigma\nu=\sigma_{\Mult}$. \hfill $\square$
\end{prop1}

\textit{Proof:}
        See Appendix~\ref{appendix Lemma 2}. 
\hfill $\blacksquare$

It is highlighted that the newborn error under the automorphism $\Phi_\theta$, which is $\Delta$ in \eqref{lemma 2 expression}, has a constant bound regardless of the error accumulated in $\cc$. 

Making use of Proposition~\ref{prop:autom}, we define the algorithm $\UnpackCt_k:R_q^2 \rightarrow R_q^{2k}$ that homomorphically evaluates $\UnpackPt_k$, which are described in Algorithms~\ref{alg:UnpackCt} and~\ref{alg:UnpackPt}, respectively. 
Observe that polynomial addition and automorphism $\Psi_\theta$ have been substituted by $\oplus$ and $\Phi_\theta$, respectively, and that the set of automorphism keys $\ak_\theta$ is necessary to execute $\UnpackCt_k$.

The remainder of the subsection serves to study the growth of errors caused by the packing and unpacking algorithms. 
Note that the external product $\boxdot$ is executed ($\log_2 \tau$)-times (one for each $\Phi_\theta$) in obtaining each $\cc_i$ of Algorithm~\ref{alg:UnpackCt}. 
This accounts for the bound on $\Delta$ stated in the following lemma.

\begin{lem1}\label{lem:UnpackProp}\upshape
    For $\cc\in R_q^{2}$ and $k\in\N$ such that $k\le \tau$, let $\UnpackCt_k(\cc)=[\cc_0;\dots;\cc_{k-1}]$ and $\Dec(\cc) = \sum_{j=0}^{N-1}\rmc_{j}X^j$. Then, for each $i=0,\dots,k-1$,
        $$\UnpackPt_k(\Dec(\cc_i)) = \begin{bmatrix}
            \rmc_{iN/\tau} \\ 0_{(k-1) \times 1}
        \end{bmatrix} + \Delta \!\! \mod q$$
        for some $\Delta \in \Z_q^{k}$ such that $\|\Delta \| \le \log_2 \tau\cdot\sigma_{\Mult}$. \hfill $\square$
\end{lem1}

\textit{Proof:}
    See Appendix~\ref{appendix lemma 3}. 
\hfill $\blacksquare$

           

The implication of Lemma~\ref{lem:UnpackProp} is that $\UnpackCt_k$ distributes the coefficients in the packing slots of $\Dec(\cc)$ into the constant terms of $\Dec(\cc_i)$. 
In particular, the newborn errors in the packing slots of $\Dec(\cc_i)$ are guaranteed to have a constant bound regardless of the accumulated error in $\cc$. 

The following proposition extends the result of Lemma~\ref{lem:UnpackProp} and states a bound on the growth of errors when homomorphically evaluating the matrix-vector multiplication \eqref{eq:matvecAB}.

\begin{prop1}\label{prop:UnpackErr}\upshape
    For $\cc\in R_q^2$ and $A\in \Z_q^{h \times l}$ such that $h\le \tau$ and $l\le \tau$, let $ \UnpackCt_l(\cc)= [\cc_0;\cdots;\cc_{l-1}]$ and $A_i$ denote the $(i+1)$-th column of $A$ for $i=0,1,\ldots,l-1$. 
    Then,
     \begin{align}\label{eq:prop2equality}
          &\UnpackPt_h(\Dec( \textstyle\sum_{i=0}^{l-1}\Enc'(\Pack_h(A_i))\boxdot \cc_i ) )\nonumber \\
            &= A \cdot \UnpackPt_l(\Dec(\cc )) + \Delta \!\! \mod q
     \end{align}
    for some $\Delta \!\in\! \mathbb{Z}_q^h$ such that $\|\Delta \| \!\le\! l(1\!+\!\| A^\top\|\log_2 \tau)\!\cdot\!\sigma_{\Mult}$.   \hfill $\square$  
\end{prop1}

\textit{Proof:}
    See Appendix~\ref{appendix proposition 2}. 
\hfill $\blacksquare$

\subsection{Encrypted controller design}\label{Sec:HoistedEnccontroller}

In this subsection, we propose an encrypted controller that integrates the introduced packing and unpacking algorithms, resulting in fewer operations needed at each time step compared to \eqref{eq:encController}.  
With a slight abuse of notation, we denote the state, input, output, and initial state of the encrypted controller by $\xx(t)$, $\yy(t)$, $\uu(t)$, and $\xx^\ini$, analogous to \eqref{eq:encController}. 

Suppose that the automorphism keys $\ak_\theta$ for $\theta \in \{2^\xi+1 \mid 2 \le 2^\xi \le  \tau, \ \xi\in\N \}$ are given, which are necessary to execute $\UnpackCt_k$.
During the offline procedure (before the control begins), the matrices $F$, $G$, and $ H$, and the vector $ x^\ini$ of \eqref{eq:nominalController} are encrypted as follows; let us denote the $(i+1)$-th column of $F$, $G$, and $ H$ by $F_i$, $ G_i$, and $ H_i$, respectively.
Each column of $F$, $G$, and $H$ is quantized, packed, and encrypted as
    \begin{align*}
    \FF_i \!&:=\! \Enc'(\Pack_n(F_i\!\!\!\mod q))\!\in\! R_q^{2 \times 2d}, \!\!&\!\! i&=0,\ldots,n-1, \\
        \GG_i \!&:=\! \Enc'(\Pack_n(G_i/\sss \!\!\!\mod q))\!\in\! R_q^{2 \times 2d}, \!\!&\!\! i&=0,\ldots,p-1,\\
        \HH_i \!&:=\! \Enc'(\Pack_m(H_i/\sss \!\!\!\mod q))\!\in\! R_q^{2 \times 2d}, \!\!&\!\! i&=0,\ldots,n-1,  
    \end{align*}
and the vector $x^\ini$ is quantized, packed, and encrypted as 
\begin{equation*}
    \xx^\ini := \Enc(\Pack_n(x^\ini/(\rrr\sss\LLL) \!\! \mod q)) \in R_q^2
\end{equation*}
using the scale factors $1/\sss$ in \eqref{eq:quantizedParams} and $1/\LLL\in\N$.

At each time step $t\in\Z_{\ge 0}$ of the online procedure, the plant output $y(t)$ is quantized, packed, and encrypted at the sensor, as 
\begin{equation}\label{eq:encController2PlantInput}
    \yy(t) := \Enc(\Pack_p(\lceil y(t)/\rrr \rfloor / \LLL \!\! \mod q))\in R_q^2,
\end{equation}
where $\rrr>0$ is the quantization step size in \eqref{eq:ypreprocess}.
It is then sent to the encrypted controller, written by
\begin{subequations}\label{eq:encController2}
\begin{align}
        \!\!\xx(t+1) \!&= \!\left(\textstyle\sum_{i=0}^{n-1} \FF_i\boxdot\xx_i(t) \!\right) \!\oplus\! \left( \textstyle\sum_{i=0}^{p-1}\GG_i\boxdot\yy_i(t) \!\right) , \label{eq:encController2State} \\
        \uu(t) \!&=\! \textstyle\sum_{i=0}^{n-1} \HH_i\boxdot\xx_i(t), \label{eq:encController2Output}\\
        \xx(0) \!&=\! \xx^\ini, \label{eq:encController2Initial}
\end{align}
where the summation is taken with respect to $\oplus$, and
\begin{equation}
    \begin{split}
        [\xx_0(t);\cdots;\xx_{n-1}(t)]&:=\UnpackCt_{n}(\xx(t)), \\
        [\yy_0(t);\cdots;\yy_{p-1}(t)]&:=\UnpackCt_{p}(\yy(t)).
    \end{split}
\end{equation}
The output $\uu(t)$ is transmitted to the actuator, where it is decrypted, unpacked, and then scaled down to obtain the plant input $u(t)$, as 
\begin{equation}
    u(t) = \rrr\sss^2\LLL \cdot \UnpackPt_m (\Dec(\uu(t))).
\end{equation}
\end{subequations}

Table~\ref{table:numOfOperations2} compares the encrypted controllers \eqref{eq:encController} and \eqref{eq:encController2} proposed in Sections~\ref{Sec:NaiveEnccontroller} and~\ref{Sec:HoistedEnccontroller}, respectively, in terms of the number of operations performed at each time step.

\subsection{Performance analysis}
 
As done in Section~\ref{Sec:Naive}, a controller over $\mathbb{R}$ whose performance is equivalent to that of the encrypted controller \eqref{eq:encController2} is derived. 
For consistency, let us abuse notation and define $\rmu_0(t)$ and $\rmx_0(t)$ (distinct from \eqref{eq:decStateOutput}), by 
\begin{equation}\label{eq:decStateOutput2}
    \begin{split}
        \rmu_0(t) &:= \rrr\sss^2\LLL \cdot \UnpackPt_m (\Dec(\uu(t)))\in\R^m,\\
        \rmx_0(t) &:= \rrr\sss\LLL\cdot \UnpackPt_n(\Dec(\xx(t)) )\in\R^n .
    \end{split}
\end{equation}
Then, we define the perturbations $\rme_0^x(t)$, $\rme_0^u(t)$, and $\rme_0^{\ini}$ by
\begin{align}\label{eq:perturbationsDef2}
        \rme_0^x(t) &:= \rmx_0(t+1) - (F\rmx_0(t) + Gy(t)), \ 
         \rme_0^{\ini} := \rmx_0(0)-x^\ini, \nonumber \\
        \rme_0^u(t) &:= \rmu_0(t)-H\rmx_0(t).
\end{align}
It is easily verified that the performance of \eqref{eq:encController2} is equivalent to that of the following controller over $\mathbb{R}$ with $u(t) = \rmu_0(t)$:
\begin{align}\label{eq:constantController2}
         \rmx_0(t+1)&=F\rmx_0(t)+Gy(t)+\rme_0^x(t), \quad \rmx_0(0)=x^{\ini}+\rme_0^{\ini},\nonumber \\
        \rmu_0(t)&=H\rmx_0(t)+\rme_0^u(t).
\end{align}

The following lemma states that the perturbations of \eqref{eq:constantController2}
remain bounded for all $t\in \Z_{\ge 0}$ if the modulus $q$ is chosen sufficiently large. 
To state the lemma, we slightly adapt $\alpha$, $\beta$, and $\bar{\eta}$ defined in Lemma~\ref{lemma1}, as
\begin{align}\label{def alpha' beta'}
        \alpha'(\rrr,\sss,\LLL)&:=\alpha(\rrr,\sss,\LLL)\!+\!\rrr\sss\LLL \!\left(\!n\!\left\|F^\top\right\| \!+\! \frac{p\left\|G^\top \right\|}{\sss}\!\right)\!\log_2 \tau\!\cdot\! \sigma_{\Mult}, \nonumber \\
        \beta'(\rrr,\sss,\LLL)&:=\beta(\rrr,\sss,\LLL)+\rrr\sss\LLL n\left\| H^\top\right\|\log_2\tau\cdot\sigma_{\Mult}, \nonumber \\
        \bar{\eta}'(\rrr,\sss,\LLL) &:= \eta(\alpha'(\rrr,\sss,\LLL),\beta'(\rrr,\sss,\LLL),\gamma(\rrr,\sss,\LLL)).
\end{align}

\begin{table}[t] 
    \caption{Number of Operations Performed at Each Time Step} 
    \centering      
    \renewcommand{\arraystretch}{1.2}
        \begin{tabular}{c|| c c } 
        \hline
          & Sec~\ref{Sec:NaiveEnccontroller}& Sec~\ref{Sec:HoistedEnccontroller} 
          \\  \hline               
        $\Enc$ & $p$ & $1$ \\ 
         \hline       
         $\Dec$ & $m$ & $1$ \\ 
           \hline  
        $\oplus$ & $n^2+n(p+m-1)-m$ & $2n+p-2$ \\ 
           \hline 
         $\boxdot$ & $n^2+n(p+m)$ & $2n+p$ \\ 
           \hline  
         $\UnpackCt_k$ & - & 2 \\ 
         \hline 
         $\UnpackPt_k$ & - & 1 \\ 
         \hline
         $\Pack_k$ & - & 1 \\ 
        \hline 
        \end{tabular} 
    \label{table:numOfOperations2} 
\end{table}

\begin{lem1}\label{lemma4}\upshape
    Given the parameters $\rrr>0$, $1/\sss\in\N$, and $1/\LLL\in\N$, the perturbations of \eqref{eq:constantController2} remain bounded by 
    \begin{equation*}
        \| \rme_0^x(t) \| \!\le\! \alpha'(\rrr,\sss,\LLL), \ \ 
        \| \rme_0^u(t) \| \!\le\! \beta'(\rrr,\sss,\LLL), \ \
        \| \rme_0^{\ini} \| \!\le\! \gamma(\rrr,\sss,\LLL),
    \end{equation*}
    for all $t\in \Z_{\ge 0}$ if 
    \begin{equation}\label{eq:lem4qbound}
         q > 2\max \left\{\frac{\bar{\eta}'(\rrr,\sss,\LLL)}{\rrr\sss\LLL}, \frac{\| H \|\bar{\eta}'(\rrr,\sss,\LLL)+  \beta'(\rrr,\sss,\LLL)}{\rrr\sss^2\LLL}   \right\}.
    \end{equation}
    \hfill $\square$
\end{lem1}
   
\textit{Proof:}
    See Appendix~\ref{appendix lemma 4}. 
\hfill $\blacksquare$

Note that Lemma~\ref{lemma4} is a generalization of Lemma~\ref{lemma1} because $\alpha' = \alpha$ and $\beta' = \beta$, and $\bar{\eta}' = \bar{\eta}$ when $\tau = 1$, where the constant term is the only packing slot. 
It is immediate from Lemma~\ref{lemma4} that the performance error in \eqref{eq:problem} can be made arbitrarily small with the choice of parameters $\{q,\rrr,\sss,\LLL \}$.

\begin{thm1}\label{theorem2}\upshape
     For given $\epsilon>0$, if the parameters $\rrr>0$, $1/\sss\in\N$, and $1/\LLL\in\N$ satisfy
     \begin{align}\label{eq:thm2Bound}
        \!\!\beta'(\rrr,\sss,\LLL ) \!\le\! \frac{\epsilon}{2} \ \ \ \mbox{and}  \ \ \
            \bar{\eta}'(\rrr,\sss,\LLL) \!\le\!  M\left\|  \begin{bmatrix}
        x_{p}^{\ini} \\
            x^{\ini}
    \end{bmatrix} \right\| \!+\! \frac{\epsilon}{2\left\|H \right\|},  
    \end{align}
    and the modulus $q$ satisfies \eqref{eq:lem4qbound}, then the encrypted controller \eqref{eq:encController2} guarantees that \eqref{eq:problem} holds for all $t \in \Z_{\ge 0}$. \hfill $\square$
\end{thm1}

\textit{Sketch of Proof:}
        The proof is similar to that of Theorem~\ref{theorem1}. Simply replace the controller \eqref{eq:constantController} and the functions $\alpha$, $\beta$, and $\bar{\eta}$ in the proof of Theorem~\ref{theorem1} with the controller \eqref{eq:constantController2} and the functions $\alpha'$, $\beta'$, and $\bar{\eta}'$, respectively.
\hfill $\blacksquare$

Since $\alpha'(\rrr,\sss,\LLL)$ and $\beta'(\rrr,\sss,\LLL)$ vanish at the origin as with $\gamma(\rrr,\sss,\LLL)$, the sufficient condition \eqref{eq:thm2Bound} can be easily satisfied by choosing sufficiently small $\rrr$, $\sss$, and $\LLL$.

\begin{rem1}\label{rem:computation}\upshape
    We provide some comparisons of the encrypted controllers \eqref{eq:encController} and \eqref{eq:encController2}.
    For the implementation of \eqref{eq:encController}, one needs to store $(n^2+np+nm)$-Ring-GSW ciphertexts in memory which correspond to the encryptions of each component of the matrices $F$, $G$, and $H$. 
    In this respect, \eqref{eq:encController2} is more memory efficient because each column of those matrices is packed before encryption, allowing one to store only $(2n+p)$-Ring-GSW ciphertexts.
    In terms of computational efficiency, it should be first noted that $\UnpackCt_k$ takes $\tau-1$ external products (one for each automorphism $\Phi_{i+1}$) for execution, and that the computation complexity of the external product is dominant over other operations such as $\oplus$ or $\Bitreverse$. 
    Therefore, \eqref{eq:encController2} is expected to take less computation time because it requires fewer external products at each time step according to Table~\ref{table:numOfOperations2}.
    This is well supported by the simulation results; see Section~\ref{Sec:Simulation}.\hfill $\square$
\end{rem1}

\section{Simulation Results}\label{Sec:Simulation}

In this section, simulation results of the proposed methods applied to the four tank system \cite{Johansson 00} are provided.
The model of the form \eqref{eq:plant} is obtained as
\begin{equation*}
    \begin{split}
        A\!&=\!\!{\footnotesize\begin{bmatrix}
            0.9984 &\! 0 &\! 0.0042 &\! 0   \\
		0 &\! 0.9989 &\! 0 &\! -0.0033  \\
		0 &\! 0 &\! 0.9958 &\! 0 \\
		0 &\! 0  &\! 0  &\! 0.9967
        \end{bmatrix}}\!, \ 
        B \!=\!\! {\footnotesize\begin{bmatrix}
             0.0083 &\! 0 \\
             0 &\! 0.0063 \\
             0 &\! 0.0048 \\
             0.0031 &\! 0
         \end{bmatrix}}\!, \\ 
        C \!&=\!\! {\footnotesize\begin{bmatrix}
            0.5 & 0 & 0  & 0      \\
            0 & 0.5 & 0  & 0
              \end{bmatrix}},
    \end{split}
\end{equation*}
by discretizing the system from \cite{Johansson 00} with the sampling time of $\SI{100}{\milli\s}$.
An observer-based controller that stabilizes the plant is designed as
\begin{equation}\label{eq:simObsvContrller}
        z(t+1) = (A+BK-LC)z(t) + Ly(t), \ \
        u(t)   = Kz(t), 
\end{equation}
where the gains $K\in\R^{2\times 4}$ and $L\in\R^{4\times 2}$ are given by 
\begin{equation*}
    \begin{split} 
        K& = {\footnotesize\begin{bmatrix}
             -0.7905 & 0.1579 & -0.2745 & -0.2686 \\
            -0.1552 & -0.7874 & -0.3427 & 0.3137
        \end{bmatrix}},  \\
        L & = {\footnotesize\begin{bmatrix}
            0.7815 & 0 & 0.3190 & 0 \\
            0 & 0.7816 & 0 & -0.3199
        \end{bmatrix}}^\top. 
    \end{split}
\end{equation*} 

Next, the state matrix of \eqref{eq:simObsvContrller} is converted to integers. 
For the sake of simplicity, we use the method of \cite[Lemma~1]{Kim 23}, which utilizes re-encryption;
let the state dynamics of \eqref{eq:simObsvContrller} be rewritten as 
\begin{align*}
        z(t+1)= (A+BK-LC-RK)z(t) + 
              \begin{bmatrix}
                L & R    
              \end{bmatrix}
              \begin{bmatrix}
                  y(t) \\ u(t)
              \end{bmatrix},
\end{align*}
where $u(t)$ is regarded as a fed-back input, and
the matrix $R\in\R^{4\times 2}$ is designed as
\begin{equation*}
     R = {\footnotesize  \begin{bmatrix}
      -1.6879 & 0.4148 & 0.2880 & -0.7385 \\
      -0.6892 & -2.1054 & 4.1931 & -2.0807
    \end{bmatrix}^\top}, 
\end{equation*}
so that $\mathrm{det}(\lambda I_4 - (A+BK-LC-RK) )=\lambda(\lambda-1)(\lambda+1)(\lambda-2)$.
Then, a coordinate transformation $x(t)=Tz(t)$
yields a controller of the form $\eqref{eq:nominalController}$ with 
\begin{equation*}
    \begin{split}
        F&=T(A+BK-LC-RK)T^{-1}\in\Z^{4\times 4} \\
        G &= T\begin{bmatrix}
            L & R
        \end{bmatrix}, \quad H = KT^{-1},
    \end{split}
\end{equation*}
which is in the modal canonical form.
The initial values are chosen as $x_p^\ini=[1;1;1;1]$ and $x^\ini=[0.5;0.02;-1;0.9]$, so that $\|x_p^\ini; x^\ini \| = 1$.

The performance of the encrypted controllers \eqref{eq:encController} and \eqref{eq:encController2} are compared in terms of computation time and performance error. 
For the simulation, i) the error distribution $\psi$ of \eqref{eq:RLWEerr} is set as the discrete Gaussian distribution with standard deviation $3.2$ that is bounded by $\sigma=6\times3.2=19.2$; ii) the modulus $q$ and the decomposition base $\nu$ of the Ring-LWE based scheme are set as $q=72057594037948417\approx 2^{56}$ and $\nu = 2^7$.
We developed our code with Lattigo version $6.1.0$ \cite{Lattigo} and all experiments are performed on a desktop with Intel Core i$7$-$12700$K CPU running $20$ threads at $3.60$ GHz.

\begin{figure}[t]
             \input{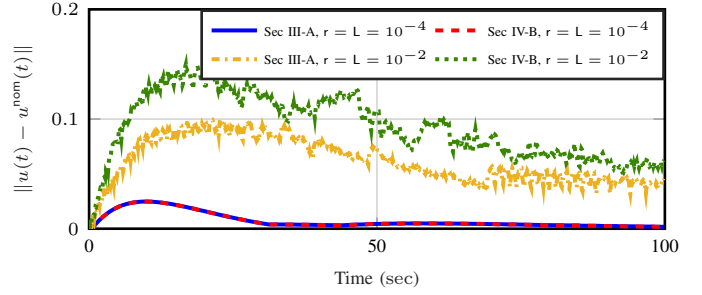}
              \caption{Performance error $\|u(t)-u^\nom(t)\|$ of the encrypted controllers \eqref{eq:encController} and \eqref{eq:encController2} for different values of $\rrr$ and $\LLL$.} 
              \label{fig:simulation}        
\end{figure}

Fig.~\ref{fig:simulation} depicts the performance error $\|u(t)-u^\nom(t)\|$ for different values of $\rrr$ and $\LLL$, when $N=2^{12}$.
It is observed that the performance error remains bounded in all cases, which asserts that the proposed encrypted controllers can operate for an infinite time horizon without bootstrapping\footnote{The selected sets of parameters $\{q,\rrr,\sss,\LLL\}$ satisfy the conditions \eqref{eq:lem1qbound} and \eqref{eq:lem4qbound}. 
The lower bounds of $\epsilon$ such that \eqref{eq:thm1Bound} and \eqref{eq:thm2Bound} both hold are computed as $1.2439\times 10^4$ and $1.2439\times 10^8$ for the cases $\rrr=\LLL=10^{-4}$ and $\rrr=\LLL=10^{-2}$, respectively. 
Nevertheless, the performance error remains below $0.2$ for $100$ seconds (1000 iterations) in all cases.}.
Moreover, it is seen that reducing the parameters $\rrr$ and $\LLL$, which can be interpreted as reducing the effects of quantization and errors injected during encryption, leads to better performance.

Table~\ref{table:compTime} shows the mean, maximum, minimum, and standard deviation of the computation time taken at each time step, when $N=2^{11}$ and $N=2^{12}$ which ensure $64$-bit and $128$-bit security \cite{AlbrPlay15}, respectively.   
The parameters $\rrr$ and $\LLL$ are set as $\rrr=\LLL=10^{-4}$.
It can be seen that the packing algorithm notably improves the computational efficiency, and that there exists a trade-off between computation time and security level.
In all cases, the computation time falls within the sampling time, suggesting the practical applicability of the proposed methods for real-time control systems.


\section{Conclusions}\label{Sec:Conclusion}
In this paper, we have introduced an encrypted dynamic controller that executes an unlimited number of recursive homomorphic multiplications on a Ring-LWE based cryptosystem without bootstrapping.
By deriving a controller over $\R$ whose performance is equivalent to that of the encrypted controller, we have shown that the effect of error growth can be arbitrarily bounded by means of the closed-loop stability. 
We have established a sufficient condition on the parameters $\{q,\rrr,\sss,\LLL\}$ under which the proposed encrypted controller guarantees a desired control performance.
Furthermore, we have proposed and applied a novel packing algorithm that leads to improved computational and memory efficiency.

\begin{table}[t] 
    \caption{ Computation time at each time step } 
    \centering      
    \renewcommand{\arraystretch}{1.2}
        \begin{tabular}{c|| c c c c c}  
        \hline
        $N$ & Method & Mean (\SI{}{\milli\second}) & Max (\SI{}{\milli\second}) & Min (\SI{}{\milli\second}) & Std \\  
        \hline               
        \multirow{2}{*}{$2^{11}$} & Sec~\ref{Sec:NaiveEnccontroller} & $12.63$ & $15.01$ & $11.00$ & $0.69$ \\
          \cline{2-6} 
        & Sec~\ref{Sec:HoistedEnccontroller} & 
        $5.59$ & $9.58$ & $3.99$ & $0.56$ 
        \\   \hline       
        \multirow{2}{*}{$2^{12}$} & Sec~\ref{Sec:NaiveEnccontroller} & $26.37$ & $31.78$ & $23.98$ & $1.06$ \\
          \cline{2-6} 
        &  Sec~\ref{Sec:HoistedEnccontroller} & 
        $11.53$ & $15.50$ & $10.44$ & $0.62$ \\  
        \hline
        \end{tabular} 
        \begin{tablenotes}
            \footnotesize 
            \item $\ast$ Number of algorithms performed at each iteration are specified in Table~\ref{table:numOfOperations2} ($n=4$ and $m=p=2$).
            \item $\ast\ast$ Parameters $\rrr=\LLL=10^{-4}$ are used.
        \end{tablenotes}
    \label{table:compTime} 
\end{table}

\appendix

\subsection{Proof of Lemma~\ref{lemma1}}\label{appendix Lemma 1}
We start by specifying the perturbations $\rme_0^x(t)$, $\rme_0^u(t)$, and $\rme_0^{\ini}$. It follows from Proposition~\ref{prop:homoProperties} that
\begin{equation*}
    \begin{split}
        &\frac{G}{\sss} \Dec(\yy(t))\!\!\mod q = \frac{G}{\sss}\left(\frac{1}{\LLL}\left\lceil \frac{y(t)}{\rrr} \right\rfloor +\Delta(t) \right) \!\!\mod q\\
        &= \frac{Gy(t)}{\rrr\sss\LLL} +\frac{G}{\sss\LLL}\left(\left\lceil \frac{y(t)}{\rrr} \right\rfloor-\frac{y(t)}{\rrr} \right) +\frac{G}{\sss}\Delta(t) \!\!\mod q
    \end{split}
\end{equation*}
for some $\Delta(t)\in R_q^p$ bounded by $\| \Delta(t) \| \le \sigma$. 
Then, the followings are immediate from Proposition~\ref{prop:homoProperties} and \eqref{eq:multHomoMat}:
\begin{align}\label{eq:lem1deltaDef}
    \frac{\rmx(t+1)}{\rrr\sss\LLL} &= \Dec((\FF \boxdot\xx(t))\oplus(\GG \boxdot\yy(t))) \\
        &=\frac{F\rmx(t)+Gy(t)}{\rrr\sss\LLL} + \Delta^x(t) \!\!\!\mod q, \nonumber \\
        \frac{\rmu(t)}{\rrr\sss^2\LLL} &= \Dec(\HH\boxdot \xx(t) ) = \frac{H\rmx(t)}{\rrr\sss^2\LLL} + \Delta^u(t) \!\!\!\mod q ,\nonumber\\
        \frac{\rmx(0)}{\rrr\sss\LLL} &= \Dec(\xx^{\ini}) = \frac{x^{\ini}}{\rrr\sss\LLL}+\Delta^{\ini} \!\!\!\mod q ,\nonumber
\end{align}
where $\Delta^x(t) \!\in\! R_q^n\!$, $\Delta^u(t) \!\in\! R_q^m\!$, and $\Delta^{\ini} \!\in\! R_q^n\!$ are bounded by
\begin{equation}\label{eq:lem1deltaBound}
    \begin{split}
        \!\!\!\!\| \Delta^x(t)  \| &\le (n+p ) \sigma_{\Mult} +\frac{\| G \|}{2\sss\LLL}+ \frac{\| G \|}{\sss} \sigma =\frac{\alpha(\rrr,\sss,\LLL)}{\rrr\sss\LLL}, \\
        \!\!\!\!\| \Delta^u(t)  \| &\le n \sigma_{\Mult}\!=\!\frac{\beta(\rrr,\sss,\LLL)}{\rrr\sss^2\LLL}, \ \
        \|\Delta^{\ini}  \| \le \sigma\!=\!\frac{\gamma(\rrr,\sss,\LLL)}{\rrr\sss\LLL}.
    \end{split}
\end{equation}
for all $t\in \Z_{\ge 0}$.

As with \eqref{eq:decStateOutput}, let $\Delta^x_i(t) \in \mathbb{Z}_q^n$, $\Delta^u_i(t) \in \mathbb{Z}_q^m$, and $\Delta^{\ini}_i \in\mathbb{Z}_q^n$ denote the vectors consisting of the $i$-th coefficients of $\Delta^x(t)$, $\Delta^u(t)$, and $\Delta^{\ini}$, respectively. 
Then, using \eqref{eq:perturbationsDef} and \eqref{eq:lem1deltaDef}, the perturbations $\rme_0^x(t)$, $\rme_0^u(t)$, and $\rme_0^\ini$ can be expressed as
\begin{equation}\label{eq:lem1perturbationsSpecify}
    \begin{split}
         \rme_0^x(t) &= \rrr\sss\LLL\cdot\left(\frac{F\rmx_0(t)+Gy(t)}{\rrr\sss\LLL}+\Delta^x_0(t) \!\!\!\mod q  \right) \\ 
          &\quad -(F\rmx_0(t)+Gy(t)), \\
           \rme_0^u(t) &= \rrr\sss^2\LLL\cdot\left(\frac{H\rmx_0(t)}{\rrr\sss^2\LLL}+\Delta^u_0(t) \!\!\!\mod q  \right)  
          -H\rmx_0(t), \\
          \rme_0^{\ini} &= \rrr\sss\LLL\cdot\left(\frac{x^{\ini}}{\rrr\sss\LLL}+\Delta^{\ini}_0 \!\!\!\mod q  \right)  
          -x^{\ini}.
    \end{split}
\end{equation}

We now show that the modulo operations in \eqref{eq:lem1perturbationsSpecify} can be omitted. 
Let us consider an auxiliary dynamics given by
\begin{equation}\label{eq:lem1virtual}
      \begin{split}
       z(t+1)&=Fz(t)+Gy(t)+\rrr\sss\LLL \cdot \Delta^x_0(t), \\
       v(t)&=Hz(t)+\rrr\sss^2\LLL \cdot \Delta^u_0(t), \\
        z(0)&=x^{\ini}+\rrr\sss\LLL \cdot \Delta^{\ini}_0,
    \end{split}
\end{equation}
where $z(t)\in\mathbb{R}^n$ and $v(t)\in\mathbb{R}^m$ are the state and the output, respectively, and $y(t)$ is the plant output of the closed-loop system \eqref{eq:plant} with \eqref{eq:lem1virtual}, where $u(t)=v(t)$. 
We claim that 
\begin{align}\label{eq:lem1claim}
    \left\| z(t) \right\|< \rrr\sss\LLL \frac{q}{2} 
 \quad \mbox{and} \quad \left\| v(t)\right\| < \rrr\sss^2\LLL\frac{q}{2}
\end{align}
hold for all $t\in \Z_{\ge 0}$. With $\chi(t) := [x_p(t);z(t)]$, the closed-loop system can be written by
     \begin{equation*}
         \begin{split}
             \chi(t+1) \!=\! \bar{A}\chi(t) + \begin{bmatrix}
             \rrr\sss^2\LLL \!\cdot\! B\Delta^u_0(t) \\ \rrr\sss\LLL \!\cdot\! \Delta^x_0(t)
         \end{bmatrix}\!\!, \,
            \chi(0)  \!=\! \begin{bmatrix}
                x_{p}^{\ini} \\
                x^{\ini} +\rrr\sss\LLL \!\cdot\! \Delta_0^{\ini}
            \end{bmatrix}\!\!.
         \end{split}
     \end{equation*}
     It follows from \eqref{eq:lem1deltaBound} and the definitions of $M$ and $\lambda$ that
     \begin{align}\label{eq:lem1boundProof}
            \left\| z(t) \right\| &\le \left\|\chi(t) \right\|  \\
            \!&= \! \left\| \bar{A}^t\chi(0)+\textstyle\sum_{k=0}^{t-1}\bar{A}^k\begin{bmatrix}
                \rrr\sss^2\LLL \cdot B\Delta^u_0(t-1-k) \\\rrr\sss\LLL \cdot \Delta^x_0(t-1-k)
            \end{bmatrix} \right\| \nonumber \\
             \!& \le\! \left\| \bar{A}^t \right\| \! \cdot \!\left\| \chi(0)\right\| \!+\!\! (\| B \| \beta(\rrr,\sss,\LLL) \!+\! \alpha(\rrr,\sss,\LLL)\! ) \!\! \textstyle\sum_{k=0}^{t-1}\! \left\| \bar{A}^k\! \right\|\nonumber \\
             \!& \le\! M\!\left( \left\|\! \begin{bmatrix}
            x_{p}^{\ini} \\
            x^{\ini} 
        \end{bmatrix} \!\right\|\!+\!\gamma(\rrr,\sss,\LLL) \!+\! \frac{ \left\| B \right\| \beta(\rrr,\sss,\LLL) \!+\! \alpha(\rrr,\sss,\LLL)}{1-\lambda} \! \right) \nonumber\\
            \!&=\! \bar{\eta}(\rrr,\sss,\LLL), \nonumber\\
              \left\| v(t) \right\| \!&\le\!  \left\| H \right\| \cdot\left\|z(t)\right\|+ \rrr\sss^2\LLL\left\|\Delta^u_0(t)\right\| \nonumber \\
              \!&\le\! \left\| H \right\| \bar{\eta}(\rrr,\sss,\LLL)+\beta(\rrr,\sss,\LLL).\nonumber
    \end{align}
    Hence, the claim holds by \eqref{eq:lem1qbound} and \eqref{eq:lem1boundProof}.
    
    Given that the claim \eqref{eq:lem1claim} holds, we can rewrite \eqref{eq:lem1virtual} as
    \begin{equation*}
        \begin{split}
           z(t+1)&=\rrr\sss\LLL\cdot \left(\frac{Fz(t)+Gy(t)}{\rrr\sss\LLL}+\Delta^x_0(t)\!\!\!\mod q \right), \\
            v(t) & = \rrr\sss^2\LLL\cdot\left(\frac{Hz(t)}{\rrr\sss^2\LLL} + \Delta^u_0(t) \!\!\! \mod q \right), \\
            z(0) &= \rrr\sss\LLL\cdot\left( \frac{x^\ini}{\rrr\sss\LLL} + \Delta^\ini_0 \!\!\! \mod q \right)
        \end{split}
    \end{equation*}
because $a=a \!\! \mod q$ for all $a\in\Z$ such that $\left | a \right | < q/2$. 
Plugging \eqref{eq:lem1perturbationsSpecify} into \eqref{eq:constantController} leads to $\rmx_0(t) \equiv z(t)$ and $\rmu_0(t) \equiv v(t)$. Therefore, the perturbations can be further specified as
\begin{align*}
    \rme_0^x(t)&=\rrr\sss\LLL \cdot \Delta^x_0(t), \!&\! 
    \rme_0^u(t)&=\rrr\sss^2\LLL\cdot\Delta^u_0(t), \!&\! 
    \rme_0^{\ini}&=\rrr\sss\LLL\cdot\Delta^{\ini}_0,
\end{align*}
and this concludes the proof combined with \eqref{eq:lem1deltaBound}.

\subsection{Proof of Theorem~\ref{theorem1}}\label{appendix Theorem 1} 

Let us denote the states of the plant and the controller of the closed-loop system \eqref{eq:plant} with \eqref{eq:nominalController} by $x_p^\nom(t)$ and $x^\nom(t)$, respectively.
Considering \eqref{eq:closedLoop}, we define the error variable, as 
      \begin{equation*}
          x_e(t):= 
          \begin{bmatrix}
              x_p(t)-x_p^\nom(t) \\
              \rmx_0(t)-x^\nom(t)
          \end{bmatrix},
      \end{equation*}
      which leads us to the error dynamics, written by 
        \begin{equation*}
                  x_e(t+1) = \bar{A}x_e(t)+ \begin{bmatrix}
                 B & 0 \\
                 0 & I_n
             \end{bmatrix}\begin{bmatrix} 
                            \rme^u_0(t)  \\ 
                            \rme^x_0(t) 
            \end{bmatrix}, \quad
             x_e(0) = \begin{bmatrix}
                 0 \\ \rme_0^{\ini}
             \end{bmatrix}.
        \end{equation*}
        Since $\| u(t)- u^\nom(t)\|= \| H(\rmx_0(t)-x^\nom(t))+\rme_0^u(t)\|$, it suffices to show that $x_e(t)$ and $\rme_0^u(t)$ are bounded by
        \begin{align}\label{eq:thm1ToShow}
              \left \| x_e(t) \right\| \le \frac{\epsilon}{2\left\| H \right\|} \quad \mbox{and} \quad  \left\|\rme_0^u(t)\right\|\le \frac{\epsilon}{2}
        \end{align}
        for all $t\in \Z_{\ge 0}$.

        In fact, the second condition of \eqref{eq:thm1ToShow} follows from \eqref{eq:lem1Show} and \eqref{eq:thm1Bound}.
        Regarding the first condition, we can derive a bound on $\|x_e(t)\|$ based on the error dynamics as done for \eqref{eq:lem1boundProof}:
        \begin{equation*}
            \begin{split}
                &\left\|x_e(t)\right\| \\ 
                &\le \left\| \bar{A}^t \right\|\left\| x_e(0)\right\| + \left(\left\| B\right \| \beta(\rrr,\sss,\LLL) + \alpha(\rrr,\sss,\LLL) \right)\textstyle\sum_{k=0}^{t-1} \left\| \bar{A}^k \right\| \\
                &\le M\left(\gamma(\rrr,\sss,\LLL) + \frac{(\left\| B\right \| \beta(\rrr,\sss,\LLL) + \alpha(\rrr,\sss,\LLL) )}{1-\lambda} \right) \\
                &= \bar{\eta}(\rrr,\sss,\LLL)- M\left\|  \begin{bmatrix}
                    x_{p}^{\ini} \\
                        x^{\ini}
                \end{bmatrix} \right\| \le \frac{\epsilon}{2\left\| H \right\|},
            \end{split}
        \end{equation*}
        where the last inequality follows from \eqref{eq:thm1Bound}. 
        This concludes the proof.
\hfill $\blacksquare$

\subsection{Proof of Proposition~\ref{prop:autom}}\label{appendix Lemma 2}
    It follows from the definitions of $\Phi_\theta$, $\ak_\theta$, and $\boxdot$ that
    \begin{equation*}
        \begin{split}
        &\Dec(\Phi_\theta(\cc,\ak_\theta))=\Dec( [
            \Psi_\theta(\rmb) ; 0] - \mathrm{ak}_\theta \boxdot [
            \Psi_\theta(\rma) ; 0] ) \\
        &= \Dec( [\Psi_\theta(\rmb) ; 0] -  (\Psi_\theta(\sk )\cdot \mathcal{G} + \mathcal{L}) \Decomp( [\Psi_\theta(\rma) ; 0])) \\
        &=\Dec([\Psi_\theta(\rmb)-\Psi_\theta(\sk)\cdot \Psi_\theta(\rma) ; 0]) -\Dec(\mathcal{L} \Decomp( [\Psi_\theta(\rma) ; 0]) ) \\
        &= \Psi_\theta(\Dec(\cc )) - \Dec(\mathcal{L} \Decomp( [\Psi_\theta(\rma) ; 0]) ), 
        \end{split}
    \end{equation*} 
    where the last equality comes from the fact that $\Psi_\theta(\rmb)-\Psi_\theta(\sk)\cdot \Psi_\theta(\rma)=\Psi_\theta(\rmb-\sk\cdot \rma) = \Psi_\theta(\Dec(\cc )).$
     As done in the proof of Proposition~\ref{prop:multErr}, it can be derived that 
     $\| \Dec(\mathcal{L} \Decomp([\Psi_\theta(\rma) ; 0]) ) \|\le dN\sigma \nu = \sigma_{\Mult},$
    which concludes the proof.

\subsection{Proof of Lemma~\ref{lem:UnpackProp}}\label{appendix lemma 3}
For convenience, let us first introduce an operation $\Slot:R_q \ra R_q$ that takes the packing slots of a given polynomial $\rmm = \sum_{i=0}^{N-1}\rmm_iX^i\in R_q$ as follows:
\begin{equation}\label{eq:defSlot}
    \Slot(\rmm) := \textstyle\sum_{i=0}^{\tau-1} \rmm_{iN/\tau}X^{iN/\tau}.
\end{equation}
Some properties of the operation $\Slot$ are provided.
    \begin{enumerate}
        \item [P1)] For any $\rmm\in R_q$ and odd $\theta \in \N$, $ \Slot(\Psi_\theta(\rmm)) = \Psi_\theta(\Slot(\rmm))$.
        \item [P2)] For any $\rmm_1\in R_q$ and $\rmm_2\in R_q$, $\Slot(\rmm_1 + \rmm_2) = \Slot(\rmm_1) + \Slot(\rmm_2)$.
    \end{enumerate}
    P1 holds because $N$ is a power of two, and thus, is coprime with $\theta$. 
    This implies that the non-packing slots do not affect the packing slots under the automorphism $\Psi_\theta$.
    P2 is trivial.

Without loss of generality, we only prove for the case of $i=0$.
Let us denote $\cc_0$ computed at Step~\ref{line8} of Algorithm~\ref{alg:UnpackCt} when $\zeta=\tau/2^{l-1}$ by $\cc_0^l$. When $l=1$ (or $\zeta=\tau$), $\cc_0^1$ is obtained as 
    \begin{equation*}\label{proposition 2 exp 1}
        \cc_0^1 = \frac{q+1}{2}\cdot\cc \oplus \Phi_{\tau+1}( \frac{q+1}{2}\cdot\cc, \ak_{\tau+1}).
    \end{equation*}
    By Propositions~\ref{prop:homoProperties} and~\ref{prop:autom}, we have
    \begin{equation*}
        \begin{split}
            \Dec(\cc_0^1 ) 
            &= \frac{q+1}{2}\cdot\Dec(\cc) + \Dec( \Phi_{\tau+1} ( \frac{q+1}{2}\cdot\cc,\mathrm{ak}_{\tau+1} ) ) \\
            &=\frac{q+1}{2}\cdot\left(\Dec(\cc) + \Psi_{\tau+1}(\Dec(\cc))\right) + \Lambda_1
        \end{split}
    \end{equation*}
    for some $\Lambda_1\in R_q$ such that $\|\Lambda_1 \|\le \sigma_{\Mult}$.
    Then, by P1-P2,
    \begin{align}\label{eq:lem3proofProp}
        &\Slot(\Dec(\cc_0^1)) \\
            &= \frac{q+1}{2} \cdot
            \left( \Slot(\Dec(\cc)) + \Slot(\Psi_{\tau+1}(\Dec(\cc))) \right) + \Slot(\Lambda_1) \nonumber \\
            & = \frac{q+1}{2} \cdot
            \left( \Slot(\Dec(\cc)) + \Psi_{\tau+1}(\Slot(\Dec(\cc))) \right) + \Slot(\Lambda_1) \nonumber\\
            &= \rmc_0 \!+\! \rmc_{2N/\tau}X^{2N/\tau} + \!\cdots\! + \rmc_{(\tau-2)N/\tau}X^{(\tau-2)N/\tau} \!+\! \Slot(\Lambda_1).\nonumber
    \end{align}
 
    When $l=2$ (or $\zeta=\tau/2$), $\cc_0^2$ is obtained from $\cc_0^1$, as
    \begin{equation*}
        \cc_0^2 = \frac{q+1}{2}\cdot\cc_0^1 \oplus \Phi_{\tau/2+1}( \frac{q+1}{2}\cdot\cc_0^1, \mathrm{ak}_{\tau/2+1}),
    \end{equation*}
    and analogous to \eqref{eq:lem3proofProp}, it can be derived that
    \begin{multline*}
        \Slot(\Dec(\cc_0^2)) = \rmc_0 + \rmc_{4N/\tau}X^{4N/\tau} + \cdots \\
             + \frac{q+1}{2}\cdot\left(\Slot(\Lambda_1) + \Psi_{\tau/2+1}(\Slot(\Lambda_1)) \right) + \Slot(\Lambda_2)
    \end{multline*}
    for some $\Lambda_2\in R_q$ such that $\|\Lambda_2 \|\le \sigma_{\Mult}$. Since $(q+1)/2$ is the multiplicative inverse of two in $\mathbb{Z}_q$ and $\|\Lambda_1 \|\le \sigma_{\Mult}$, 
    \begin{equation*}
        \left\|\frac{q+1}{2}\!\cdot\!\left(\Slot(\Lambda_1) \!+\! \Psi_{\tau/2+1}(\Slot(\Lambda_1)) \right) \!+\! \Slot(\Lambda_2) \right\| \!\le\! 2\sigma_{\Mult}.
    \end{equation*}
   Applying the same argument repeatedly until $l=\log_2 \tau$ (or $\zeta=2$), we arrive at
   \begin{equation*}
       \Slot(\Dec(\cc_0^{\log_2\tau})) = \rmc_0 + \Slot(\bar\Lambda)
   \end{equation*}
   for some $\bar\Lambda\in R_q$ such that $\|\Slot(\bar\Lambda)\|\le \log_2\tau \cdot \sigma_{\Mult}$, and this concludes the proof.

\subsection{Proof of Proposition~\ref{prop:UnpackErr}}\label{appendix proposition 2}

    Some additional properties of the operation $\Slot$ defined in \eqref{eq:defSlot} are first provided.
     \begin{enumerate}
        \item [P3)] For any $\rmm_1\in R_q$ and $\rmm_2\in R_q$, $\Slot(\Slot(\rmm_1)\cdot \rmm_2)= \Slot(\rmm_1)\cdot \Slot(\rmm_2)$.
        \item [P4)] For any $a\in\Z_q^k$ and $k\in\N$ such that $k\le \tau$, $\Slot(\Pack_k(a)) = \Pack_k(a)$.
         \item [P5)] For any $a\in\Z_q^k$, $\rmm\in R_q$, and $k\in\N$ such that $k\le \tau$, $\|\Pack_k(a) \cdot \Slot(\rmm)\| \le \left\|a^\top\right\|\cdot\|\Slot(\rmm)\|$.
        \item [P6)] For any $\rmm\in R_q$ and $k\in\N$ such that $k\le \tau$, $\UnpackPt_k(\Slot(\rmm)) =\UnpackPt_k(\rmm) $.
    \end{enumerate}
    P3 and P5 follow from the definition of polynomial multiplication in \eqref{eq:defAddMult}, and the fact that the packing slots are equidistant. 
    P4 and P6 are trivial. 
    
    By Proposition~\ref{prop:multErr} and P2-P4, it holds that
    \begin{equation*}
        \begin{split}
        &\Slot(\Dec( \textstyle\sum_{i=0}^{l-1}\Enc'(\Pack_h(A_i))\boxdot \cc_i) ) \\
        &= \Slot(\textstyle\sum_{i=0}^{l-1}\Pack_h(A_i)\cdot \Dec(\cc_i) + \Lambda) \\ 
        &= \textstyle\sum_{i=0}^{l-1} \Slot(\Pack_h(A_i)\cdot \Dec(\cc_i)) + \Slot(\Lambda) \\
        & = \textstyle\sum_{i=0}^{l-1} \Slot(\Slot(\Pack_h(A_i))\cdot \Dec(\cc_i)) + \Slot(\Lambda) \\
        & =  \textstyle\sum_{i=0}^{l-1} \Pack_h(A_i)\cdot \Slot(\Dec(\cc_i)) + \Slot(\Lambda)=: \Omega \in R_q 
        \end{split}
    \end{equation*}
    for some $\Lambda \in R_q$ such that $\left\|\Lambda \right\| \le l\cdot\sigma_{\Mult}$. 
     Let $\Dec(\cc) = \sum_{j=0}^{N-1}\rmc_jX^{j}$.
    By Lemma~\ref{lem:UnpackProp}, $\Slot(\Dec(\cc_i)) = \rmc_{iN/\tau} + \Slot(\Lambda_i)$ for some $\Lambda_i\in R_q$ such that $\| \Slot(\Lambda_i) \| \le \log_2 \tau \cdot \sigma_\Mult$ for each $i=0,1,\ldots,l-1$. 
    Hence,
    \begin{equation*}
        \begin{split}
            \Omega\!&=\! \textstyle\sum_{i=0}^{l-1} \Pack_h(A_i)\!\cdot\! \rmc_{iN/\tau} + \Pack_h(A_i)\!\cdot \!\Slot(\Lambda_i) +\Slot(\Lambda)\\
            \!&=: \!\textstyle\sum_{i=0}^{l-1} \Pack_h(A_i)\cdot \rmc_{iN/\tau} + \bar\Lambda,
        \end{split}
    \end{equation*}
    where $\bar\Lambda\!\in\! R_q$ is bounded as $\| \bar\Lambda \| \!\le\! l(1+\|A^\top\|\log_2\tau)\!\cdot\!\sigma_\Mult$ by P5.
    Thanks to P6, the left-hand side of \eqref{eq:prop2equality} is equivalent to $\UnpackPt_h(\Omega)$. This concludes the proof because $\UnpackPt_h(\textstyle\sum_{i=0}^{l-1} \Pack_h(A_i)\cdot \rmc_{iN/\tau}) \!=\! A \cdot \UnpackPt_l(\Dec(\cc )) $
    and $\| \UnpackPt_h(\bar \Lambda) \| \le \| \bar\Lambda\|$.

\subsection{Proof of Lemma~\ref{lemma4}}\label{appendix lemma 4}

    Applying Proposition~\ref{prop:UnpackErr} to \eqref{eq:encController2State} yields 
        \begin{equation*}
            \begin{split}
                &\frac{\rmx_0(t+1)}{\rrr\sss\LLL}= \UnpackPt_n (\Dec(\xx(t+1))) \\
                &= \frac{F\rmx_0(t)}{\rrr\sss\LLL} \!+\! \frac{G}{\sss}  \UnpackPt_p(\Dec(\yy(t))) \!+\! \Delta^x_1(t) \!\!\! \mod q,
            \end{split}
        \end{equation*}
        for some $\Delta^x_1(t) \in \mathbb{Z}_q^n$ such that 
        \begin{equation*}
                 \|\Delta^x_1(t)\| \!\le\! \left(\!(n+p)\!+\!\left(\!n\left\|F^\top \right\|+\frac{p\left\| G^\top\right\|}{\sss} \right)\log_2\tau\!\right)\cdot\sigma_{\Mult}.
        \end{equation*} 
        Since $\lceil y(t)/\rrr \rfloor/\LLL = y(t)/(\rrr\LLL) + (\lceil y(t)/\rrr \rfloor/\LLL - y(t)/(\rrr\LLL))$,
        \begin{equation*} 
                 \frac{G}{\sss}\UnpackPt_p(\Dec(\yy(t))) = \frac{Gy(t)}{\rrr\sss\LLL} + \Delta^x_2(t) \!\!\! \mod q
        \end{equation*}
        for some $\Delta^x_2(t)\in \mathbb{Z}_q^n$ such that 
        \begin{equation*}
           \left\|\Delta^x_2(t) \right\| \le \frac{\|G \|}{2\sss\LLL} + \frac{\|G \|}{\sss}\sigma.
        \end{equation*}
        Then, with $\Delta^x(t) := \Delta_1^x(t) + \Delta_2^x(t)$, we have
        \begin{equation*}
            \frac{\rmx_0(t+1)}{\rrr\sss\LLL} = \frac{F\rmx_0(t) + Gy(t)}{\rrr\sss\LLL} + \Delta^x(t) \!\!\! \mod q,
        \end{equation*}
        and $\|\Delta^x(t) \| \le \|\Delta^x_1(t) \| + \|\Delta^x_2(t) \| \le \alpha'(\rrr,\sss,\LLL)/\rrr\sss\LLL$.
        Similarly, applying Proposition~\ref{prop:UnpackErr} to \eqref{eq:encController2Output} and \eqref{eq:encController2Initial} results in
        \begin{equation*}
            \begin{split}
                \frac{\rmu_0(t)}{\rrr\sss^2\LLL} 
                &= \frac{H}{\sss}  \UnpackPt_n(\Dec(\xx(t))) + \Delta^u(t) \!\!\! \mod q \\
                &=\frac{H\rmx_0(t)}{\rrr\sss^2\LLL} + \Delta^u(t) \!\!\! \mod q \\
                \frac{\rmx_0(0)}{\rrr\sss\LLL} &= \frac{x^\ini}{\rrr\sss\LLL} + \Delta^\ini \!\!\! \mod q
            \end{split}
        \end{equation*}
         for some $\Delta^u(t)\in \mathbb{Z}_q^m$ and $\Delta^\ini \in \Z_q^n$ such that 
         \begin{equation*}
             \left\| \Delta^u(t) \right\| \le n\left(1+\frac{\left\| H^\top\right\|}{\sss}\log_2 \tau \right)\cdot\sigma_{\Mult} = \frac{\beta'(\rrr,\sss,\LLL)}{\rrr\sss^2\LLL}
         \end{equation*}
         and $\| \Delta^\ini \| \le \gamma(\rrr,\sss,\LLL)/\rrr\sss\LLL$.
         From this point, the remainder of the proof is parallel to that of Lemma~\ref{lemma1}, and is thus omitted.


\begin{IEEEbiography}[{\includegraphics[width=1in,height=1.25in,clip,keepaspectratio]{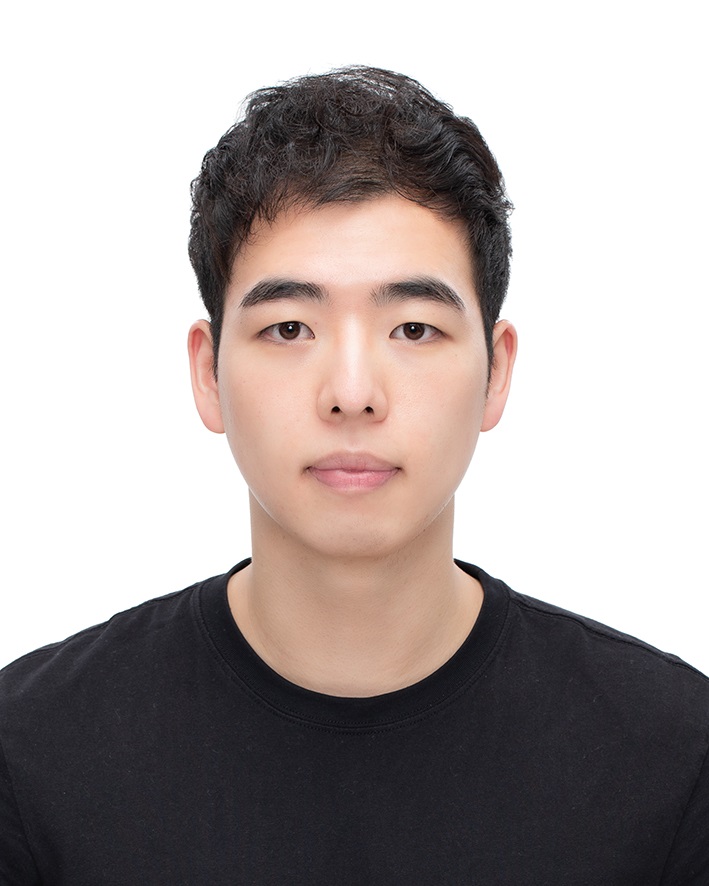}}]%
{Yeongjun Jang}
received the B.S. degree in electrical and computer engineering in 2022, from Seoul National University, South Korea.
He is currently a combined M.S./Ph.D. student in electrical and computer engineering at Seoul National University, South Korea. 
His research interests include data-driven control and encrypted control systems.
\end{IEEEbiography}

\vskip -2.5\baselineskip plus -1fil

\begin{IEEEbiography}[{\includegraphics[width=1in,height=1.25in,clip,keepaspectratio]{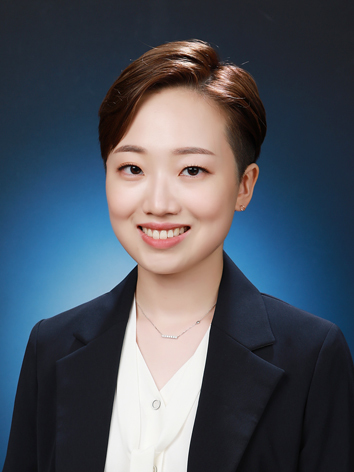}}]%
{Joowon Lee}
received the B.S. degree in electrical and computer engineering in 2019, from Seoul National University, South Korea.
She is currently a combined M.S./Ph.D. student in electrical and computer engineering at Seoul National University, South Korea. 
Her research interests include data-driven control, encrypted control systems, and cyber-physical systems.
\end{IEEEbiography}

\vskip -2.5\baselineskip plus -1fil

\begin{IEEEbiography}[{\includegraphics[width=1in,height=1.25in,clip,keepaspectratio]{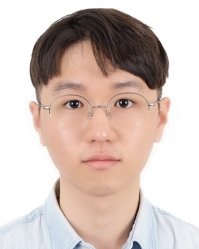}}]%
{Seonhong Min}
received the B.S. degree in mathematical sciences in 2022, from Seoul National University, South Korea.
He is currently a combined M.S./Ph.D. in computer science and engineering at Seoul National University, South Korea. 
His research interests include fully homomorphic encryption and its applications, such as privacy preserving machine learning and multi-party computation.
\end{IEEEbiography}

\vskip -2.5\baselineskip plus -1fil

\begin{IEEEbiography}[{\includegraphics[width=1in,height=1.25in,clip,keepaspectratio]{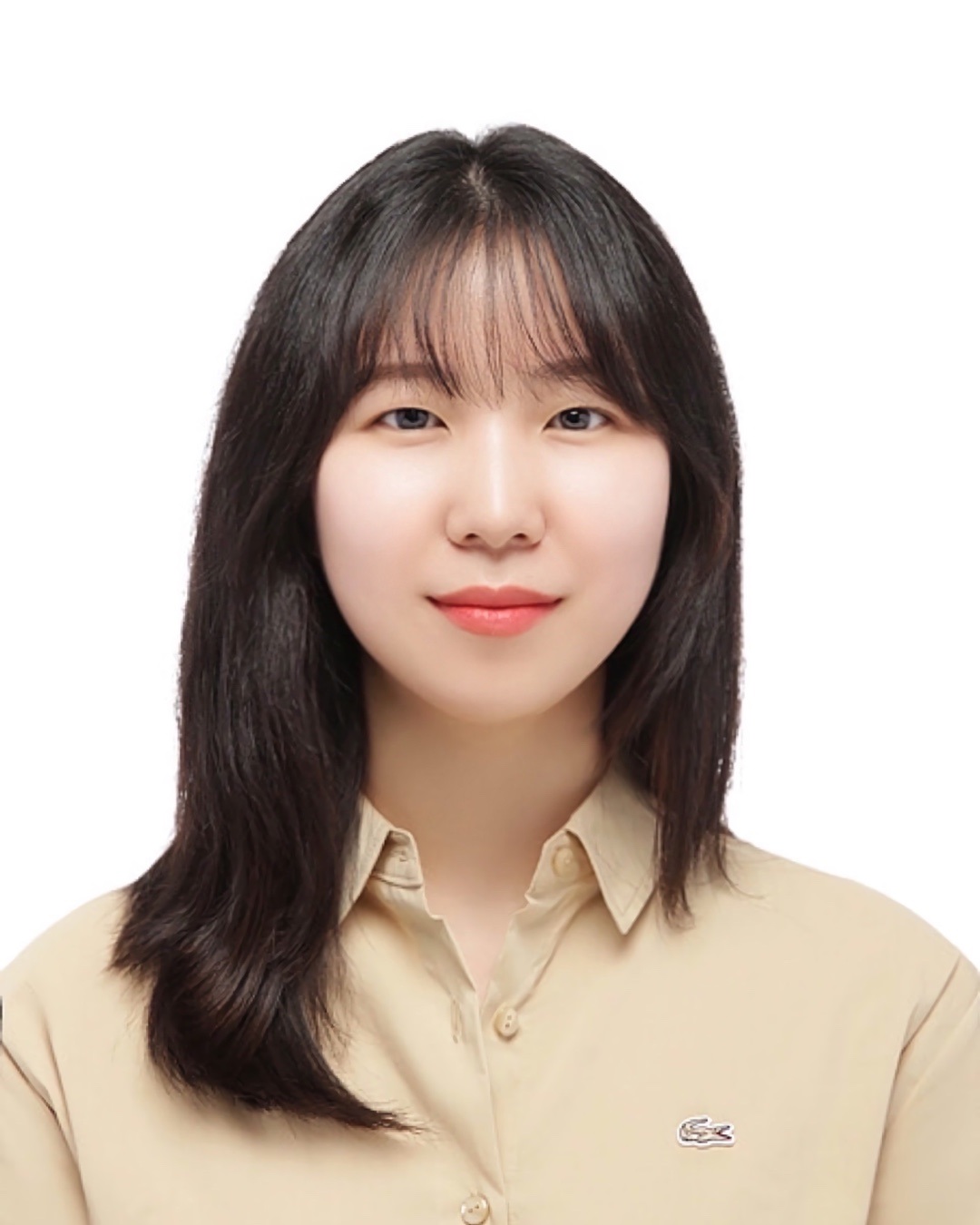}}]%
{Hyesun Kwak}
received the B.S. degrees in mathematics education and computer science and engineering in 2021, and the M.S. degree in computer science and engineering in 2023, from Seoul National University, South Korea, respectively.
She is currently a software engineer in Samsung Electronics, South Korea.
Her research interests include homomorphic encryption and its applications, such as privacy-preserving machine learning.
\end{IEEEbiography}

\vskip -2.5\baselineskip plus -1fil

\begin{IEEEbiography}[{\includegraphics[width=1in,height=1.25in,clip,keepaspectratio]{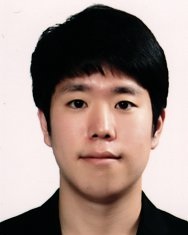}}]%
{Junsoo Kim}
received the B.S. degrees in electrical engineering and mathematical sciences in
 2014, and the M.S. and Ph.D. degrees in electrical engineering in 2020, from Seoul National University, South Korea, respectively. 
 He held the Postdoc position at KTH Royal Institute of Technology, Sweden, till 2022. 
  He is currently an
 Assistant Professor at the Department of Electrical and Information Engineering, Seoul National University of Science and Technology, South Korea. His research interests include security
 problems in networked control systems and encrypted control systems.
\end{IEEEbiography}

\vskip -2.5\baselineskip plus -1fil

\begin{IEEEbiography}[{\includegraphics[width=1in,height=1.25in,clip,keepaspectratio]{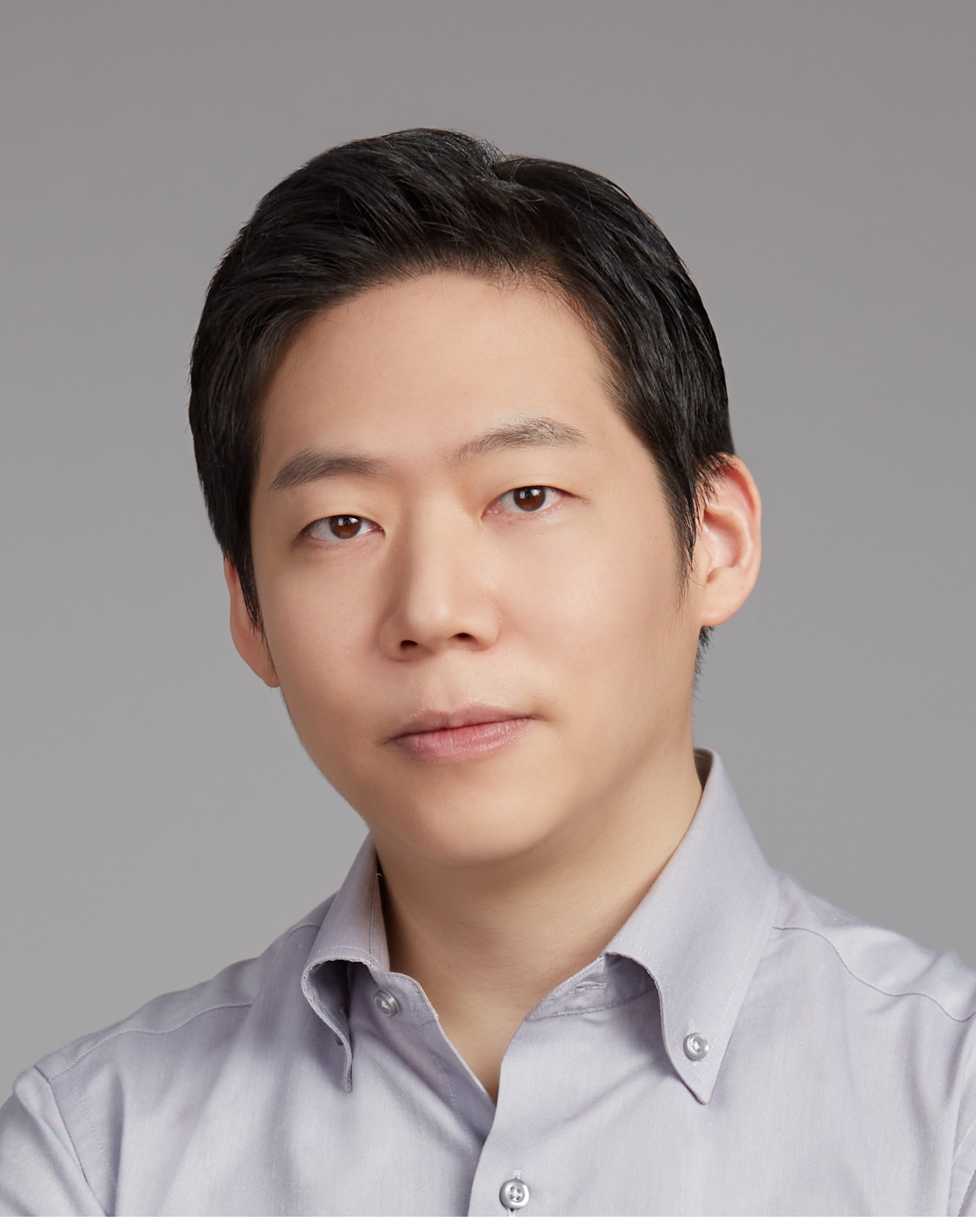}}]%
{Yongsoo Song}
received the B.S. degree in mathematical sciences in 2012, and the M.S. and Ph.D. degrees in mathematical sciences in 2018, from Seoul National University, South Korea, respectively. He held the Postdoc position at UC San Diego, USA, till 2019. 
From 2019 to 2021, he was a senior researcher in Cryptography and Privacy Research Group at Microsoft Research Redmond, USA. 
He is currently an assistant professor at the Department of Computer Science and Engineering, Seoul National University, South Korea.
His research specializes in privacy-enhancing cryptography for secure computation, which includes fully homomorphic encryption, multi-party computation, and zero-knowledge proofs, approached from both theoretical and applied perspectives.
\end{IEEEbiography}

\vfill


\begin{thebibliography}{00}
\bibitem{Sandberg 15} H.~Sandberg, S.~Amin, and K.\,H.~Johansson, ``Cyberphysical security in networked control systems: An introduction to the issue," \textit{IEEE Control Syst. Mag.}, vol.~35, no.~1, pp.~20--23, 2015.
\bibitem{Teixeira 15} A.~Teixeira, I.~Shames, H.~Sandberg, and K.\,H.~Johansson, ``A secure control framework for resource-limited adversaries," \textit{Automatica}, vol.~51, pp.~135--148, 2015.
\bibitem{Kogiso 15} K.~Kogiso and T.~Fujita, ``Cyber-security enhancement of networked control systems using homomorphic encryption," in \textit{Proc. 54th IEEE Conf. Decision Control}, 2015, pp.~6836--6843.
\bibitem{Schulze Darup 21} M.~Schulze~Darup, A.\,B.~Alexandru, D.\,E.~Quevedo, and G.\,J.~Pappas, ``Encrypted control for networked systems: An illustrative introduction and current challenges," \textit{IEEE Control Syst. Mag.}, vol.~41, no.~3, pp.~58--78, 2021.
\bibitem{ARC} J.~Kim, D.~Kim, Y.~Song, H.~Shim, H.~Sandberg, and K.\,H.~Johansson, ``Comparison of encrypted control approaches and tutorial on dynamic systems using learning with errors-based homomorphic encryption,'' \textit{Annu. Rev. Control}, vol.~54, pp.~200--218, 2022.
\bibitem{Regev 09} O. Regev, ``On lattices, learning with errors, random linear codes, and cryptography,'' \textit{J. ACM}, vol.~56, no.~6, 2009, Art. no.~34.
\bibitem{Gentry 09} C.~Gentry, ``Fully homomorphic encryption using ideal lattices," in \textit{Proc. 41st Annu. ACM Symp. Theory Comput.}, 2009, pp.~169--178.
\bibitem{Kim 23} J.~Kim, H.~Shim, and K.~Han, ``Dynamic controller that operates over homomorphically encrypted data for infinite time horizon," \textit{IEEE Trans. Autom. Control}, vol.~68, no.~2, pp.~660--672, 2023.
%
\bibitem{Gentry 13} C.~Gentry, A.~Sahai, and B.~Waters, ``Homomorphic encryption from learning with errors: Conceptually-simpler, asymptotically-faster, attribute-based," in \textit{Annu. Cryptol. Conf.}, 2013, pp.~75--92.
\bibitem{Lyubashevsky 13} V.~Lyubashevsky, C.~Peikert, and O.~Regev, ``On ideal lattices and learning with errors over rings," \textit{J. ACM}, vol.~60, no.~6, 2013, Art. no.~43.
\bibitem{Microsoft Seal} ``Microsoft SEAL (release 4.1),'' Jan. 2023, Microsoft Research, Redmond, WA. [Online]. Available:  https://github.com/Microsoft/SEAL
\bibitem{HElib} S.~Halevi, and V.~Shoup, ``Design and implementation of HElib: A homomorphic encryption library,'' \textit{Cryptol. ePrint Arch.}, Paper 2020/1481, 2020. [Online]. Available: https://eprint.iacr.org/2020/1481
\bibitem{Lattigo} ``Lattigo v6,'' Aug. 2024,
EPFL-LDS, Tune Insight, SA. [Online]. Available: https://github.com/tuneinsight/lattigo
\bibitem{Teranishi23} K.~Teranishi, T.~Sadamoto, and K.~Kogiso, ``Input-output history feedback controller for encrypted control with leveled fully homomorphic encryption,'' \textit{IEEE Trans. Control Netw. Syst.}, vol.~11, no.~1, pp.~271--283, 2023.
\bibitem{Lee 24} J.~Lee, D.~Lee, J.~Kim, and H.~Shim, ``Encrypted dynamic control exploiting limited number of multiplications and a method using RLWE based cryptosystem,'' \textit{IEEE Trans. Syst. Man Cybern.: Syst.}, early access, Oct. 23, 2024, doi: 10.1109/TSMC.2024.3452007.
\bibitem{Chillotti 16} I.~Chillotti, N.~Gama, M.~Georgieva, and M.~Izabach\'{e}ne, ``Faster fully homomorphic encryption: Bootstrapping in less than $0.1$ seconds," in \textit{Proc. Int. Conf. Theory Appl. Cryptol. Inf. Secur.}, 2016, pp.~3--33.
\bibitem{Cheon 18} J.\,H.~Cheon, K.~Han, H.~Kim, J.~Kim, and H.~Shim, ``Need for controllers having integer coefficients in homomorphically encrypted dynamic system,''
in \textit{Proc. 57th IEEE Conf. Decision Control}, 2018, pp.~5020--5025.
\bibitem{Kim 21} J.~Kim, H.~Shim, H.~Sandberg, and K.\,H.~Johansson, ``Method for running dynamic systems over encrypted data for infinite time horizon without bootstrapping and re-encryption," in \textit{Proc. 60th IEEE Conf. Decision Control}, 2021, pp.~5614--5619.
\bibitem{Tavazoei 23} M.\,S.~Tavazoei, ``Non-minimality of the realizations and possessing state matrices with integer elements in linear discrete-time controllers,'' \textit{IEEE Trans. Autom. Control}, vol.~68, no.~6, pp.~3698--3703, 2023.
\bibitem{Lee 23} J.~Lee, D.~Lee, S.~Lee, J.~Kim, and H.~Shim, ``Conversion of controllers to have integer state matrix for encrypted control: Non-minimal order approach,'' in \textit{Proc. 62nd IEEE Conf. Decision Control}, 2023, pp.~5091--5096.
\bibitem{Dowler 13} D.\,A.~Dowler, ``Bounding the norm of matrix powers,'' Master's thesis, Dept. Math., Brigham Young Univ., Provo, UT, USA, 2013.
\bibitem{Johansson 00} K.\,H.~Johansson, ``The quadruple-tank process: A multivariable laboratory process with an adjustable zero,'' \textit{IEEE Trans. Control Sys. Technol.}, vol.~8, no.~3, pp.~456--465, 2000.
\bibitem{AlbrPlay15} M.\,R.~Albrecht, R.~Player, and S.~Scott, ``On the concrete hardness of learning with errors,'' \textit{J. Math. Cryptol.}, vol.~9, no.~3, pp.~169--203, 2015.

\end{thebibliography}
\end{document}